\documentclass[11pt]{article}   %\pdfoutput=1
\usepackage[a4paper,hmarginratio=1:1,vmarginratio=2:3,totalwidth=15.5cm,
totalheight=21.5cm]{geometry}
\usepackage{bm,epstopdf,epsfig,graphicx,amsmath,amssymb,
amsfonts,latexsym,cancel,verbatim,comment,ulem}
\usepackage[usenames,dvipsnames]{color}

\newcommand{\cL}{{\cal L}}
\newcommand{\cO}{{\cal O}}
\newcommand{\cD}{{\cal D}}
\newcommand{\opsi}{{\overline\psi}}
\newcommand{\ra}{{\rightarrow}}

\newcommand{\beq}{\begin{equation}}
\newcommand{\eeq}{\end{equation}}
\newcommand{\bea}{\begin{eqnarray}}
\newcommand{\eea}{\end{eqnarray}}
\newcommand{\ba}{\begin{array}}
\newcommand{\ea}{\end{array}}
\newcommand{\bec}{\begin{center}}
\newcommand{\eec}{\end{center}}
\newcommand{\bei}{\begin{itemize}}
\newcommand{\eei}{\end{itemize}}
\newcommand{\nn}{\nonumber}

\long\def\symbolfootnote[#1]#2{\begingroup
\def\thefootnote{\fnsymbol{footnote}}\footnote[#1]{#2}\endgroup}

\title{}
\begin{document}

\thispagestyle{empty}

%\vspace{-3cm}

\begin{flushright}
{\footnotesize CPHT-RR051.0611, LPT-ORSAY 11-55 \\
 SACLAY-T11/063, CERN PH-TH-2011-153}
\end{flushright}

\vspace{1.5cm}

\begin{center}

{\bf\Large  On non-universal Goldstino couplings to matter}

\vspace{1cm}

{\bf E. Dudas}$^{\,a,b}$, {\bf G. Gersdorff}$^{\,a}$,
{\bf D.~M. Ghilencea}$^{\,c,\,d}$,
{\bf S. Lavignac}$^{\,e}$,
{\bf J. Parmentier}$^{\,a,e}$

%\vspace{0.7cm}
\bigskip

{\it \small
$^{a}$ Centre de Physique Th\'eorique\!
\symbolfootnote[2]{Unit\'e mixte du CNRS (UMR 7644).},
  Ecole Polytechnique and CNRS,
F-91128 Palaiseau, France.\\[3pt]
$^{b}$ Laboratoire de Physique Th\'eorique\!
\symbolfootnote[3]{Unit\'e mixte du CNRS (UMR 8627).},
  Universit\'e de Paris-Sud,    
B\^at. 210, F-91405 Orsay, France.\\ [3pt]
$^{c}$ Theory Division, CERN, 1211 Geneva 23, Switzerland. \\[3pt]
$^{d}$\,DFT, National Institute of Physics and Nuclear Engineering (IFIN-HH) Bucharest 
MG-6, Romania\\[3pt]
$^{e}$ Institut de Physique Th\'eorique\!
\symbolfootnote[4]{Laboratoire de la Direction des Sciences de la
Mati\`ere du Commissariat \`a l'Energie Atomique et Unit\'e de
Recherche associ\'ee au CNRS (URA 2306).}, CEA-Saclay, 
  F-91191 Gif-sur-Yvette Cedex, France.}
\end{center}

\medskip
\begin{abstract}
\noindent
Using the constrained superfields formalism to describe the interactions of
a light goldstino to matter fields in supersymmetric models, 
we identify generalised, higher-order holomorphic  superfield  constraints  
that project out the superpartners and capture the non-universal 
couplings of the goldstino to matter fields.
These arise from microscopic theories in which heavy superpartners masses are
of the order of the supersymmetry breaking scale ($\sqrt f$).
In the decoupling limit of infinite superpartners masses, these constraints
reduce to the familiar, lower-order universal constraints discussed recently,
that describe the  universal goldstino-matter fields couplings, suppressed by
inverse powers of $\sqrt f$. 
We initiate the study of the couplings of the Standard Model (SM) fields to
goldstino in the constrained superfields formalism.
\end{abstract}

%%\today
\newpage
\tableofcontents

\section{Introduction} 
\label{sec:intro}

The study of the interactions of the goldstino with matter is an interesting
research area, that started originally with the pioneering
work of Akulov and Volkov \cite{Volkov:1973ix},
and was investigated extensively  in the past \cite{cddfg}-\cite{Samuel:1982uh},
with renewed  interest in \cite{Casalbuoni:1988xh} -\cite{adgt}.
Such theoretical study could also be of  phenomenological relevance if the scale
of supersymmetry breaking (in the hidden sector), $\sqrt{f}$, is in the few-TeV range,
near the Large Hadron Collider (LHC) energy; however, this
work is valid independent of such restrictions for $f$.
The first formalisms describing goldstino interactions with matter, using in particular
constrained superfields, were constructed in the early days of supersymmetry
\cite{Ivanov:1977my}, \cite{Rocek:1978nb}, \cite{Lindstrom:1979kq}, \cite{Samuel:1982uh}.
More recently, an approach starting from  microscopic  descriptions of the
supersymmetric theories was further considered \cite{Casalbuoni:1988xh}, 
\cite{SK} in which heavy superpartners were integrated 
out via field equations, leading to constrained
superfields and renewed interest in this method.

An intriguing aspect of the constraints used in \cite{SK}
is that their solution is {\it unique} and are {\it insensitive} to the
details of the UV physics, in particular are independent of the superparticle masses $m_i$.
As a result, the corresponding low-energy actions contain {\it universal} couplings
of the goldstino to matter fields,  suppressed by $\sqrt f$. It
is clear however that the decoupling of heavy superpartners can also
lead to {\it non-universal couplings}, suppressed at low-energy by
their masses $m_i$ rather than the supersymmetry breaking scale $\sqrt{f}$.
The main purpose of the present paper is to clarify the emergence of the
non-universal couplings in a superfield formalism endowed  with constraints,
generated by the decoupling of the heavy superpartners. To this purpose we extend
the formalism of constrained superfields developed so far and identify
new, higher order  polynomial constraints for superfields,  that effectively project
out (integrate out) the massive superpartners of the SM spectrum.
One possible advantage of this is that one could then use, in  principle at
least, a pure superfield formalism   endowed with such  generalised constraints
to describe the non-supersymmetric  SM Lagrangian.

Our study shows that the non-universal couplings of the goldstino to matter
are described at the low-energy by  superfields with higher order, non-minimal
constraints. Rather interestingly, and unlike the universal case, the solution
to these generalised constraints is {\it not unique},
but depends on some arbitrary  coefficients. In specific microscopic models (i.e. UV
completions) these  coefficients are functions of the  superpartners masses that
ultimately depend on the parameters of the UV microscopic theory.  The
aforementioned arbitrariness of these  coefficients seems interesting
and may turn out to be an advantage.  Indeed, since the constraints (and solutions)
that we find are the same for all classes of UV completions (of different UV parameters),
they may turn out to be applicable beyond the limited valability of the convergent
effective  Lagrangian expansion in powers of $1/m_i^2$. This could allow an extension
of the  valability of the superfield formalism with generalised constraints beyond
that of the effective Lagrangian obtained upon integrating out the superpartners,
into the non-perturbative regime. In the formal limit of infinite superpartners masses
the goldstino couplings to matter become universal and the generalised, higher order
constraints of the superfields that we identify recover the
universal constraints.

When integrating out the superpartners, one should ensure that they are massive enough
to decouple at the low-energy $E$ at which the Lagrangian is studied, and light enough
for the UV and supersymmetry breaking scales separation. 
In detail, if the superpartners  masses 
$m_i$ are too light $E \lesssim m_i \ll \sqrt f$,  then one cannot integrate
them out,  at energy scales $E\sim m_i$; so the formalism 
is less reliable since the expansion in  $E/m_i$ is not rapidly convergent.
If on  the contrary,  the  superpartners are massive enough, of the order 
of supersymmetry breaking scale, $E \ll m_i \lesssim \sqrt f < \Lambda$,
and in the absence of some  cancellations of the aforementioned coefficients 
(in the microscopic action), then it means $m^2_i \approx f^2/\Lambda^2 \sim f$,
therefore $\Lambda^2\sim f$, where $\Lambda$ is the UV scale.
This last relation means  that there would be no significant scale separation between
the scales $m_i\sim \sqrt f$  and $\Lambda$,  therefore there
would be no reliable UV microscopic description. The conclusion is that one has
to work in a regime of energy values between these two ``dangerous'' choices,
and in applications this remains a difficult issue.
However,  note that  holomorphic superfield  constraints
(and their solutions) that we identify to project out massive scalar
superpartners are similar for different superpartners masses and UV physics 
details. Therefore, they could be more powerful in capturing aspects of UV physics
without a specific UV completion or a concern for scales separation,  
and, as argued earlier, could even capture effects of  non-perturbative physics.

The plan of the paper is as follows. Section~\ref{sec:nonlinear} reviews
the goldstino couplings and the universal constraints formalism.
In Section~\ref{sec:constrained} we identify generalised holomorphic superfield
constraints for generic Kahler potentials of matter fields and goldstino,
and superpotentials  with additional matter couplings.
In Section~\ref{sec:leading} we use the new  formalism to provide
examples with the R-parity violating couplings in supersymmetric model, 
in the decoupling limit of heavy squarks and sleptons masses. We extend this
and evaluate  the onshell Lagrangian of goldstino coupled to a general
Kahler potential for matter fields, with additional  superpotential
matter couplings.  
Section~\ref{sec:simplest} outlines the steps for writing
the Standard Model and its two-Higgs doublet model extension coupled
to a light goldstino  in the superfield formalism by using
universal constraints only, with particular attention to 
the Higgs sector.  We find a new quartic correction to the Higgs
potential similar to that in \cite{adgt}, but unlike that, 
it is not universal, but depends on additional parameters.
We also point out that in general, the leading higgs-goldstino couplings are not
universal, being generated by a term analogous to the $\mu$-term in
MSSM. The  Appendix  contains some comments on  the effects on the formalism of
constrained superfields from microscopic terms that contain derivatives 
of light fields, and microscopic examples generating heavy gauginos and 
higgsinos.

%%%%%%%%%%%%%%%%%%%%%%%%%%%%%%%%%%%

\section{Non-linear supersymmetry and constrained superfields}
\label{sec:nonlinear}

In this section we review the goldstino action and its (universal) couplings
to the SM, in a superfield formalism with constraints.
We consider in the following energy scales well below 
the mass of the (scalar) sgoldstino $\phi_X$, the superpartner of goldstino $\psi_X$.
We denote the goldstino chiral superfield  by $X$ with the 
components $X=(\phi_X,\psi_X,F_X)$.
 One can use the component fields formalism to describe the 
 corresponding  Volkov-Akulov action
and then couple it ``geometrically'' to the SM action.
 Alternatively, one can use the more convenient superfield formalism,
endowed with constraints. For the goldstino superfield, this constraint
is \cite{Rocek:1978nb,Lindstrom:1979kq,Casalbuoni:1988xh,Brignole:1997pe,SK}.
\beq
X^2 \ = \ 0 \ . \label{nl1}
\eeq
Starting with the Lagrangian\footnote{We use the
conventions of \cite{WB}.}
\medskip
\beq
\label{X}
{\cal L}_X\!=\!
\int\! d^4\theta\,  X^\dagger X
+\bigg\{\!\int\! d^2\theta \,f\,X+h.c.\bigg\}
= \vert \partial_\mu\phi_X\vert^2 + F_X^\dagger F_X
\!- \Big[
\frac{i}{2}\overline \psi_X \overline\sigma^\mu
\partial_\mu \psi_X\! - f F_X +\! h.c. \Big]
\eeq

\smallskip\noindent
endowed with the aforementioned constraint, one
recovers the Volkov-Akulov action. Indeed,
the constraint is solved by
\bea
\label{nl01}
X &=&\phi_X + \sqrt 2 \ \theta \, \psi_X
\ + \ \theta^2 \ F_X, \qquad{\rm with}\qquad
\phi_X=\frac{\psi_X \psi_X}{2\, F_X}
\eea
which, when used back in eq.(\ref{X})  recovers \cite{SK}
 the Volkov-Akulov Lagrangian \cite{Volkov:1973ix}. After using the equation
of motion  for $F_X$, we find
$F_X = - f+....$, where $f$ (that can be chosen real)
is the SUSY breaking scale. Therefore, in the infrared description
of the SUSY breaking, the  scalar component (sgoldstino) becomes a function of
the goldstino $\psi_X$.

Let us now review the (universal) couplings of the goldstino
to the SM matter fields, in the superfield formalism.
{Consider} a SM fermion field (quark or lepton),  denoted
$\psi_{q}$ in what follows and introduce  the corresponding superfield
$Q=(\phi_q,\psi_q,F_q)$.
We assume that the scalar (squark or slepton) has a large mass, so it decouples
at low (EW) scales.  The decoupling can be described in the superfield language,
 by imposing a constraint on the corresponding  superfield $Q$, which effectively
``projects out'' (i.e. integrates out)  $\phi_{q}$.
The simplest constraint  that realizes  this is \cite{Brignole:1997pe,SK}
\beq
Q \, X \ \ = \ 0 \ , \label{nl2}
\eeq
whose solution is
\beq
Q\ = \ \frac{1}{F_X} \Big( \psi_{q} - \frac{F_{q}}{2 F_X} \psi_X\Big)
\psi_X \ + \ \sqrt{2} \ \theta \ \psi_{q} + \theta^2 F_{q} \ .
\label{nl3}
\eeq
Solutions (\ref{nl01}) and (\ref{nl3}) of the constraints can be obtained in the
low energy limit of the following microscopic  Lagrangian that couples $X$ and $Q$:
\medskip
\bea
{\cal L} \ = \ \int d^4 \theta \left\{ X^{\dagger} X +  Q^{\dagger} \,Q -
\epsilon_x (X^{\dagger} X)^2 -
\epsilon_q (Q^{\dagger} Q) (X^{\dagger} X) \right\} +
\Big\{\int d^2 \theta \ f \ X +h.c.\Big\}
\, \label{nl4}
\eea

\medskip\noindent
where $\epsilon_{x,q} \equiv {c_{x,q}}/{\Lambda^2}$,
 $\Lambda > \sqrt{f}$ is a UV cutoff scale and the positive, dimensionless
coefficients $c_{x,q}$  are of order $c_X, c_{q} \sim \cO(1)$. The sgoldstino
and squark (slepton) masses are then:
\medskip
\begin{equation}
m_X^2 \ = \  4 \epsilon_x f^2 \quad , \quad
m_{q}^2 \ = \ \epsilon_q f^2  \ . \label{nl02}
\end{equation}

\medskip\noindent
The microscopic Lagrangian in (\ref{nl4}) has a minimal structure in the sense
that it contains the minimal number of terms needed
to provide large scalar masses and to stabilize the scalar vev's to zero, while
keeping vanishing fermions masses. After integrating out the massive scalar fields
$\phi_X$ and $\phi_{q}$ by using the {\it zero momentum part} of the 
Lagrangian, one obtains  solutions (\ref{nl01}), (\ref{nl3}), as anticipated. 
A naive procedure would be to consider zero-momentum for heavy fields but 
keep non-vanishing momenta (derivatives) for the light fields.  Such procedure 
would % not lead to
modify the  constraints in (\ref{nl1}), (\ref{nl2}) (for an example
see Appendix A).

One  notices that, somewhat intriguingly, the constraints $X^2=0$, $X\,Q=0$
and their solution in (\ref{nl01}), (\ref{nl3}) are actually independent of the UV details,
in this case the masses of the heavy scalars. However, this is not the most general case:
in a microscopic theory  {only mildly}  different from (\ref{nl4}),
other terms in the Kahler potential and superpotential can be present,
compatible with all the symmetries of the low-energy theory. As we shall see later
in detail, such terms can  change the constraints  (\ref{nl1}), (\ref{nl2}) in the
generic case $m_X^2,m_{q}^2~\sim~f$.
In the  extreme limit of  $m_X^2,m_{q}^2 \gg f$, all the goldstino couplings to
matter become universal and the constraints (\ref{nl1}), (\ref{nl2}) and their solution
are recovered.  However,  since $\Lambda^2\! > \! f$,
such {\it formal} limit  of
very heavy scalars is only possible for unnaturally
large values of the coefficients $c_x, c_{q} \gg 1$, that
question the reliability of the expansion in ($1/\Lambda$).
Even so, such large values for $c_x, c_{q}$ are difficult  to obtain, 
even in a strongly coupled regime when one would expect at most
$c_i\sim 4 \pi$.
Ultimately, the constraints $X^2=0$ and $X\,Q=0$ will receive UV dependent corrections,
however they can still be used  to identify the universal couplings
of the goldstino to matter fields, via this limit. 
Finally, the constraints we shall identify in the
following sections will help one identify the non-universal goldstino couplings,
and will recover these universal constraints as a special case.
There can be exceptions to this: 
even in the limit of large  $c_x, c_q$ etc, with some 
ratios of them kept fixed,  it is possible that the universal
constraints $X^2=X\,Q=0$ change to become nonzero in $\cO(1/f^2)$ 
order\footnote{See for example Section~\ref{goldstino} and
eq.(\ref{consk9}).}. 
However, by taking large values only for those coefficients contributing to
the physical masses, we do recover the universal constraints.

This discussion can be extended to include gauge multiplets.
The question is then what constraint to impose
 in order to integrate out the heavy gauginos. A simple choice 
is \cite{SK}
\medskip
\begin{equation}
X \ W_{\alpha} \ = \ 0 \ , \label{nl5}
\end{equation}
where $W_\alpha$ is the superymmetric field  strength.
The solution is given by
\medskip
\begin{eqnarray}
 W_{\alpha} &= & \frac{1}{\sqrt{2}F_X} (\delta_{\alpha}^{\beta} D-i
\sigma_{\alpha}^{mn,\beta} F_{mn}) 
\psi_{X,\beta} - \frac{\psi_X\psi_X}{2 F_X^2} (\sigma^m \partial_m {\overline
   \lambda})_{\alpha} +
 (\delta_{\alpha}^{\beta} D-i \sigma_{\alpha}^{mn,\beta} F_{mn})
\theta_{\beta} \nonumber \\[4pt]
& + &
\ \theta^2 \ (\sigma^m \partial_m {\bar  \lambda})_{\alpha} \  \label{nl6}  
\end{eqnarray}

\medskip\noindent
Note that in (\ref{nl6}) the gaugino is defined implicitly,
the solution can be found in Appendix, eq.(\ref{vector5}).  
As discussed in Appendix \ref{sec:consvector}, similarly to the
above case of heavy scalars,  eq.(\ref{nl5}) is exact only in the formal limit of
infinitely heavy gaugino masses, $M^2\gg f$. In the case $M^2 \sim f$
one could consider instead  a different constraint, $X \ W_{\alpha} W^{\alpha} = 0$.
However, the universal goldstino coupling to the gauge field is indeed generated
by both types of constraints.

Finally, let us discuss the Higgs sector. For the Higgs multiplet(s), 
one would like to decouple the fermions (higgsinos) 
instead of the scalars.  
The simplest choice is provided by the constraints \cite{SK}
\medskip
\begin{equation}
X \ H_i^{\dagger} \ = \ {\rm chiral} \quad \leftrightarrow \quad X \ 
{\bar D}_{\dot \alpha} H_i ^{\dagger} \ = \ 0 \ , \label{nl06}  
\end{equation}

\medskip\noindent
which determine the higgsinos and the whole Higgs superfields in terms of
the higgs scalars $h_i$ and the goldstino $\psi_X$ 
\medskip
\begin{equation}
H_i \ = \ h_i + i \sqrt{2} \ \theta \ \sigma^m \Big(\frac{ \opsi_X}{\overline F_X}\Big)
\partial_m h_i + \theta^2 \Big[ - \partial_m \Big(\frac{\opsi_X}{\overline F_X}\Big)
  {\overline \sigma}^n \sigma^m \frac{\opsi_X}{\overline F_X} \partial_n h_i +
  \frac{1}{2 {\overline F_X}^2} {\opsi_X}^2 \Box h_i  \Big] \ . 
\label{nl07}
\end{equation}

\medskip\noindent
We discuss in Appendix \ref{sec:higgsinos} a microscopic model
leading to a value of the higgsino as shown above, $\psi_{h,i}
=({i}/{\overline F_X}) \, \sigma^m { \opsi_X} \partial_m h_i $,
(although the details of the low-energy theory and
in particular the auxiliary fields $F_{h,i}$ are slightly different).

The various superfield constraints presented above raise  an interesting question.
What are the fundamental, ``primary''  superfield constraints in a given
model ? More explicitly,  consider the case of the goldstino $X$ and 
matter fields $Q_i$ discussed earlier, with the solution (\ref{nl01}), (\ref{nl3})
satisfying the constraints $X^2 = X Q_i = 0$.
By an explicit calculation  and using eq.(\ref{nl3})
one shows that the following relation ($i$ labels a particular fermion)
\medskip
\begin{equation}
Q_i \ Q_j \ Q_k \ = \ 0 \ . \label{nl60}
\end{equation}

\medskip\noindent
is also satisfied and can itself be regarded as a constraint.
At first sight this does not seem to be immediately implied by
the ``primary'' constraints (\ref{nl1}) and (\ref{nl2}), but eq.(\ref{nl60}) cannot be an
independent  one since the  constraints  $X^2 = X Q_i=0$ uniquely
determine solutions (\ref{nl01}),~(\ref{nl3}). The answer to these questions
is related to the generalized class of constraints that encode not only the universal
but also the additional, non-universal couplings  of the goldstino to the SM fields,
that we discuss in the next sections.

%%%%%%%%%%%%%%%%%%%%%%%%%%%%%%%%%%%%%%%%%%%%%%%%%%%%%%%%%%%%%%%

\section{UV versus effective Lagrangians and generalized constraints}

\label{sec:constrained}

In this section we identify new, generalised constraints for superfields
and their relation to integrating out massive superpartners.
The framework is that of supersymmetric theories % (for example the MSSM), 
in which all would-be SM superpartners 
are heavy and decoupled at low energy. In order to integrate 
out the heavy scalars\footnote{We comment in 
Appendix~\ref{sec:consvector} and \ref{sec:higgsinos} on
the case of integrating heavy fermions like higgsinos/gauginos.},
via their classical field equations, we use the
zero-momentum limit of the Lagrangian of a rigid supersymmetric theory
with $N$ fields,  given by \cite{zumino}
\medskip
\bea
\cL &=&
\int d^4\theta \, K(\Phi^\dagger_I,\Phi^I)
+\Big\{\int d^2\theta \,\, W(\Phi^I)\,+h.c.\Big\}
\nonumber\\
&=& \!\!\!
K_I^{\,\,\,J} \Big[\partial_\mu\phi^I\,\partial^\mu\phi_J^\dagger
- \frac{i}{2}\,\big( \psi^I\,\sigma^\mu\cD_\mu\opsi_J
- \cD_\mu \,\psi^I\sigma^\mu\opsi_J\big)
+ F^I\,F_J^\dagger\Big]
\nonumber\\
&+&
\frac{1}{4}\,K_{IJ}^{KL}\,\psi^I\psi^J\,\opsi_K\opsi_L
%-\frac{1}{4}\,\Box\,K(\phi^i\,\phi_j^\dagger)
+\Big[\big(W_K - \frac{1}{2} \,K_K^{IJ}\opsi_I\opsi_J\big)
\,F^K - \frac{1}{2}\,W_{IJ}\,\psi^I\psi^J
 +h.c.\Big]\label{generalKW}
\label{cons2}\eea

\medskip\noindent
where the lower (upper) indices in $K$ and $W$ functions
denote derivatives wrt scalar fields (their hermitian conjugate),
respectively, in a standard notation, $\cD_\mu \,\psi^I$ denote
the Kahler covariant derivative acting on the fermion $\psi^I$,
$K_I=\partial K/\partial \Phi^I$, $K^J=\partial K/\partial \Phi^\dagger_j$,
 etc. Here $\Phi_I=(\phi_I,\psi_I,F_I)$
can denote $X$ or a matter superfield, such as that of a quark/squark ($Q$) 
or lepton/slepton. For such a theory, the goldstino is defined as
\medskip
\bea
&& \psi_X
= \frac{1}{|F|}
 {\overline F}_{I}\, \psi^I \ = \ - \  \frac{1}{|F|} (K^{-1})_{I}^{\,J} \, W_J \,\psi^I \ ,
\qquad {\rm where} \qquad |F|^2 \ = \ \sum_I |F_I|^2 \   \label{cons1}
\eea

\medskip\noindent
is the total magnitude of supersymmetry breaking. Expression
(\ref{cons1}) is valid on the ground state, so all scalars should be
replaced by their vev's.

Integrating out the massive superpartners can be performed in terms of superfield
constraints that we identify below; their form  will depend on the
complexity of the Kahler potential  and superpotential. Nevertheless, in all
cases these constraints will be some  polynomials in  superfields. Let us provide some examples.

%%%%%%%%%%%%%%%%%%%%%%%%%%%%%%%%%%%%%%%%%%

\subsection{Goldstino in microscopic theories and universal constraints.}
\label{goldstino}

To  begin with, consider a particular  case of the microscopic
action  in (\ref{cons2}), with $N$ fields
\bea
&& K \ = \ %\sum_i
\Phi_I^{\dagger} \Phi^I \ - \  \frac{1}{\Lambda^2} \ \Big(
 \Phi_I^{\dagger} \Phi^I\Big)^2 \ , \qquad
W \ = \ \sum_I \ f_I \ \Phi_I \ ,
\label{goldstino1}
\eea
with summation over $I={\overline{1,N}}$;
here the Kahler potential preserves a global $U(N)$ symmetry while  $f_I$ are some
constants, chosen real. Using a zero-momentum integration of
the heavy scalars one finds
\medskip
\beq
\Phi_I \ = \Big[\,
\frac{1}{|F|^2} \ \Big(\psi_I - F_I \ \frac{{\overline F_J}\, \psi^J }{2 |F|^2} \Big)
\ {\overline F_N}\, \psi^N\, \Big] \ + \ \sqrt{2} \ \theta \ \psi_I
\ + \ \theta^2 \ F_I \ , \label{goldstino2}
\eeq

\medskip\noindent
where a summation over $J,N$ is understood.
It is important to notice that in this case $\Phi_I^2 \not=0$ 
 for each field $\Phi_I$. However, the superfield
\medskip
\beq
X \ = \ \frac{1}{|F|} \ {\overline F}_{J} \,\Phi^J \
 = \ \frac{({\overline F_J}\, \psi^J)^2}{2 |F|^3} \ + \ \sqrt{2} \ \theta
\ \frac{{\overline F_J}\, \psi^J}{|F|} \ + \ \theta^2 \ |F|
\eeq

\medskip\noindent
satisfies the constraint $X^2=0$, with its goldstino as defined in 
eq.(\ref{cons1}). Further, any linear combination
$Q_I = c_{IJ} \Phi^J$ can be written in the form (\ref{nl3}) with $\psi_q\ra \psi_{q_I}
= c_{IJ} \psi^J$, $F_q \ra F_{q_I} = c_{IJ} F^J$  and therefore satisfies the
constraint in (\ref{nl2}), $X Q=0$. 

However,  the details of the microscopic theories are important and  
the constraint $X^2=0$ is not valid in all such theories,  even though the 
goldstino fermion remains that defined as in (\ref{cons1}).  
Instead of the relation $X^2=0$,   a higher order monomial constraint applies, 
as we shall see shortly in more general microscopic models.
The case discussed above is rather special, due to the $U(N)$ symmetry 
of the microscopic Kahler potential (\ref{goldstino1}); this  lead
to equal masses for the heavy scalars and therefore to only one mass scale
and then  $X^2=0$ follows.

%%%%%%%%%%%%%%%%%%%%%%%%%%%%%%%%%

\subsection{Kahler potentials and generalized chiral constraints}
\label{sec:consk}

The  Kahler potential in eq.(\ref{nl4}) is the minimal one required
by a consistent decoupling of the heavy scalars, but it is not protected by 
any symmetry. Therefore, it can contain additional  terms of the form
$(Q^{\dagger} Q)^2/\Lambda^2$ in the matter fields sector. New terms in the 
Kahler potential ($K$) introduce new mass scales, and the form of 
$K$ and the total number of low-energy fields determine the nature
of the constraints on the low-energy superfields.
Let us provide an example with one goldstino superfield $X$ and one SM matter fermion
whose superfield is $Q$.   The microscopic action that we consider 
is then (we ignore gauge interactions):
\medskip
\bea
&&
K \ = \  X^{\dagger} X \ + \ Q^{\dagger} Q \ - \ \epsilon_1 (X^{\dagger} X)^2 \
- \ \epsilon_2 (Q^{\dagger} Q)^2 \ - \ \epsilon_3 (Q^{\dagger} Q)
 (X^{\dagger} X) \ , \nonumber \\[5pt]
&&
W \ = \  f \ X \ , \qquad {\rm where}\qquad Q\equiv(\phi_q,\psi_q,F_q)
\label{consk1}
\eea

\medskip\noindent
where $\epsilon_i\sim 1/\Lambda^2$ have mass dimension $-2$ and $\Lambda$ is the UV scale.
The scalar fields have squared  masses~\footnote{For $\epsilon_1,\epsilon_3>0$, 
there is a local minimum at the origin, $\langle X\rangle=0$, $\langle Q\rangle=0$; 
this is the vacuum we work with in what follows.} of values  $4 \epsilon_1 f^2$ for sgoldstino
 and $\epsilon_3 f^2$ for  the squark (or slepton), respectively. 
We integrate out these scalar fields,  using the zero-momentum Lagrangian for the 
heavy scalar fields  {\it and also}  for the light fermion fields. 
One finds the low-energy superfields
\medskip
\bea
 X & = & \big( a_{11}\,\psi_X\psi_X + a_{12}\, \psi_X \psi_q 
+ a_{22}\, \psi_q \psi_q\,\big) \
+ \ \sqrt{2} \ \theta \ \psi_X + \theta^2 F_X \ , \nonumber \\
Q & = & \big( \,b_{11} \,\psi_X\psi_X + b_{12}\, \psi_X \psi_q 
+ b_{22}\, \psi_q \psi_q\,\big) \
+ \ \sqrt{2} \ \theta \ \psi_q + \theta^2 F_q \ , \label{consk2}
\eea

\medskip\noindent
where the coefficients have the expressions
\medskip
\bea
&& a_{11} = \frac{\epsilon_1 {\overline F}_X}{\Delta} (\epsilon_3 |F_X|^2
+ 4 \epsilon_2 |F_q|^2) \ , \qquad
a_{12} = \frac{2 \epsilon_2 \epsilon_3 F_q {\overline F}_q^2}{\Delta} \ , \qquad
a_{22} = - \frac{\epsilon_2 \epsilon_3 F_X {\overline F}_q^2}{\Delta}\ ,
\nonumber \\[7pt]
&&
b_{11} = - \frac{\epsilon_1 \epsilon_3 F_q {\overline F}_X^2}{\Delta} \ , \qquad
b_{12} = \frac{2 \epsilon_1 \epsilon_3 F_X {\overline F}_X^2}{\Delta} \ , \qquad
b_{22} = \frac{\epsilon_2 {\overline F}_q}{\Delta} (\epsilon_3 |F_q|^2
+ 4 \epsilon_1 |F_X|^2),\qquad
 \label{consk3}
\eea
and
\beq
\Delta \ = \ 2 (\epsilon_1 \epsilon_3 |F_X|^4 + \epsilon_2 \epsilon_3 |F_q|^4
+ 4 \epsilon_1 \epsilon_2 |F_X|^2 |F_q|^2) \ . 
\eeq

\medskip\noindent
One  can check that the low-energy
superfields (\ref{consk2}) satisfy the following cubic constraints
\medskip
\beq
X^3 \ = \ X^2 Q \ = \ X Q^2 \ = \ Q^3 \ = \ 0 \qquad {\rm but}
\qquad
X^2\not=0 \ , \quad  X \,Q \ \not=0 \ . \label{consk4}
\eeq

\medskip\noindent
A peculiar feature of the constraints in (\ref{consk4})
is that, unlike the ``minimal'' ones in  eqs.(\ref{nl1}), (\ref{nl2}), their 
solution is not unique. The solution of (\ref{consk4})  with the ansatz (\ref{consk2}) is
\medskip
\bea
X \ &=&
\ \Big(
\frac{\psi_X\psi_X}{2 F_X} \ - \ \frac{c_1}{2 F_X} (F_q \psi_X - F_X \psi_q)^2\Big)
+\sqrt 2 \,\theta\psi_X+ \theta^2\,F_X
 \ , \nonumber \\
\quad
Q \ &=& \Big(
\ \frac{\psi_q \psi_q}{2 F_q} \ - \ \frac{c_2}{2 F_q} (F_q \psi_X - F_X \psi_q)^2\, \Big)
\, +\,\sqrt 2 \,\theta\psi_q+ \theta^2\,F_q \ 
 , \label{consk5}
\eea

\medskip\noindent
which are valid for arbitrary coefficients $c_{1,2}$.
In practice, however, $c_{1,2}$ are nontrivial functions of the auxiliary fields
and for our case, eq.(\ref{consk3}), can be written as
\medskip
\bea\label{cc}
c_1=\frac{2}{\Delta}\,\epsilon_2\epsilon_3 \,{\overline F_q}^2,\qquad
c_2=\frac{2}{\Delta}\,\epsilon_1\epsilon_3 \,{\overline F_X}^2 \ . 
\eea

\medskip\noindent
For $\epsilon_2=0$ one has $c_1=0$, $c_2=1/F_X^2$
 and one recovers from (\ref{consk5}) the familiar case of the solutions
(\ref{nl01}), (\ref{nl3}) that respect eqs.(\ref{nl1}), (\ref{nl2}).
The non-unique character  of the solutions (\ref{consk5}) is related to the fact that in
specific examples  they depend on the UV data,
i.e. the parameters of the UV Kahler potential, as seen from  eq.(\ref{cc}).

The coefficients $c_{1,2}$ which are functions of the fields
can be expanded in powers  of the ratio $F_q/F_X$. Indeed, since $F_q$ is small
($\langle F_q\rangle \sim \epsilon_i \ll \langle F_X \rangle\sim f $), the leading-order
Lagrangian in the number of fermion fields corresponds to the lowest terms
in $F_q$ and we can  write
\medskip
\bea\label{taylor}
&& c_1 \ = \ \frac{1}{F_X^2} \bigg(
\alpha_0 + \alpha_1 \frac{F_q^2}{F_X^2} + \alpha_2 \frac{\overline F_q^2}{\overline F_X^2} +
\alpha_3 \,\bigg\vert{\frac{F_q}{F_X}}\bigg\vert^2 + \cdots \bigg) \ , \nonumber \\
&&c_2 \ = \ \frac{1}{F_X^2} \bigg( \beta_0 + \beta_1 \frac{F_q^2}{F_X^2}  +
\beta_3\, \bigg\vert\frac{F_q}{F_X}\bigg\vert^2 + \cdots \bigg) \ .  \label{consk6}
\eea

\medskip\noindent
The form of this expansion (even powers of the ratios $F_q/F_X$) is due to
the symmetries of the UV Lagrangian (\ref{consk1}). Onshell this expansion is
actually finite, since $F_X$ and $F_q$ contain bilinears in $\psi_{X,q}$.
For the expansion to be well-defined, it is necessary that $\beta_2=0$ and $\beta_0=1$.
For our model (\ref{consk1}), $\alpha_0=0$, $\alpha_1=0$, $\alpha_2=\epsilon_2/\epsilon_1$,
$\beta_0=1$, $\beta_1=0$, $\beta_3=-4\epsilon_2/\epsilon_3$.
The low-energy, effective  Lagrangian corresponding  to (\ref{consk1}) contains  non-universal
interactions that depend on the ratio of parameters $\epsilon_i$ only in the
 sub-leading terms in the expansion (\ref{consk6}).
Non-universal interactions are therefore suppressed by powers of $F_q/F_X$.

Additional UV terms can be present in the Kahler potential of (\ref{consk1}). Such
terms can modify the dominant terms in the corresponding low-energy, effective
 action, obtained after integrating out the massive scalars fields.
Consider for example the addition to (\ref{consk1}) of the term
\medskip
\begin{equation}
\delta K \ = \ - \,
\epsilon_4 \ ( X^{\dagger})^2 Q^2 \ + \  {\rm h.c.}  \ .  \label{consk7}
\end{equation}

\medskip\noindent
Such term can be present in extensions of the MSSM in which $Q$ is 
not a squark/slepton anymore, but would denote a gauge singlet  or a modulus superfield
(if the modulino fermionic component   $\psi_q$ is an accidentally  light fermion), in string
theory. The constraints in (\ref{consk4}) and the parametrization (\ref{consk5})
remain valid. We  find
\medskip
\begin{eqnarray}
&& c_1 \ = \frac{2}{\Delta'}\, \epsilon_3 (\epsilon_2 {\overline F}_q^2 -
\epsilon_4 {\overline F}_X^2)  \quad , \qquad
c_2 \ = \frac{2}{\Delta'}\,\epsilon_3 (\epsilon_1 {\overline F}_X^2
 - \epsilon_4 {\overline F}_q^2) \ , \quad {\rm where } \nonumber \\[7pt]
&&  \Delta'  \ =2\Big[ \ \epsilon_3 F_X^2  (\epsilon_1 {\overline F}_X^2 
- \epsilon_4 {\overline F}_q^2)
+ \epsilon_3 F_q^2  (\epsilon_2 {\overline F}_q^2 - \epsilon_4 {\overline F}_X^2)
+ 4 (\epsilon_1 \epsilon_2 - \epsilon_4^2) |F_X|^2 |F_q|^2 \Big]\ .
\label{consk8}
\end{eqnarray}

\medskip\noindent
In this case, the explicit solution for the heavy scalars is modified even at
the zero-th order in the expansion (\ref{consk6}),
giving $c_1=(-\epsilon_4/\epsilon_1) \,(1/F_X^2)+\cO(1/F_X^4)$ 
and $c_2=1/F_X^2+\cO(1/F_X^4)$.
In this approximation the sgoldstino is changed according to
\medskip
\begin{equation}
X \ = \ \frac{1}{2 F_X} \Big(\psi_X\psi_X + \frac{\epsilon_4}{\epsilon_1}
\psi_{q} \psi_q  \Big)
   +\sqrt 2 \,\theta\psi_X+\theta^2\,F_X \ . 
\end{equation}

\medskip\noindent
The above form of the sgoldstino modifies in turn
the leading, universal goldstino couplings in the low-energy action.
One can then  see that $X^2\not=0$, since
\medskip
\begin{equation}
\phi_{X^2}=\frac{1}{2\,F_X^2}\,\frac{\epsilon_4}{\epsilon_1}
(\psi_X\psi_X)\,(\psi_q\psi_q)\not=0 \ .
\qquad 
\label{consk9}
\end{equation}

\medskip\noindent
To complete the above discussion, we replace above
the auxiliary fields by their values from the eqs of motion:
\medskip
\beq
F_X  = - f
- 2\,\epsilon_1\,\phi_X^\dagger \,\psi_X\psi_X
- \, \epsilon_3\,\phi_q^\dagger \,\psi_X\psi_q
- 2 \,\epsilon_4\,\phi_X^\dagger\,\psi_q\psi_q
- f\,\epsilon_3\,\vert\phi_q\vert^2
- 4\,f\,\epsilon_1\,\vert\phi_X\vert^2+\cO(\epsilon_i^2\,f^2)
\nonumber
\eeq
\beq\label{os}
F_q = -\Big[
 2\,\epsilon_4\,\phi_q^\dagger\,\psi_X\psi_X
+ \,\epsilon_3\,\phi_X^\dagger\,\psi_X\psi_q
+2\,\epsilon_2\,\phi_q^\dagger\,\psi_q\psi_q
+4\,f\,\epsilon_4\,\phi_q^\dagger\,\phi_X
+ f\,\epsilon_3\,\phi_q\phi_X^\dagger\Big]
+\cO(\epsilon_i^2\,f^2)
\eeq

\medskip\noindent
Using the system of coupled eqs.(\ref{os}) and  $\phi_X, \phi_q$ identified from
(\ref{consk5}), (\ref{consk8}), one finds the onshell result for
sgoldstino and squarks/sleptons  in the microscopic model of
(\ref{consk1}) with (\ref{consk7})
\medskip
\bea\label{onshellscalars}
\phi_X & = & - \frac{\psi_X\psi_X}{2\,f}
-\frac{\epsilon_4}{\epsilon_1}\,\frac{\psi_q\psi_q}{2\,f}+\cO(\epsilon_i^2\,f^2)
\ , 
\nonumber\\
\phi_q & = & -\frac{\psi_X\psi_q}{f}
\, - \,
\frac{1}{f^3}\,(\epsilon_2-\epsilon_4^2/\epsilon_1)
\,(\psi_X\psi_X)\,(\opsi_X\opsi_q)\,(\psi_q\psi_q)+\cO(\epsilon_i^2\,f^2)
\ . 
\eea

\medskip\noindent
For $\epsilon_4 =0$ one verifies that $X^2=XQ=0$, even though, as we saw earlier,
this is not valid before eliminating the auxiliary fields. 

The above analysis can be generalised to an arbitrary number $N$ of
matter fermions $Q^i$. In this case, we obtain chiral superfield constraints given by the
vanishing of all possible superfield monomials of order $N+2$
constructed out of the superfields
$X$ and $Q^i$, $i=1,N$. Notice that in order to find a monomial constraint of the type
$P(X_i) = 0$, one checks the vanishing of its
$\theta$-independent and $\theta$ component, by using Grassmann
identities of the type $\psi^3=0$, $(\psi_1 \psi_2)^2
= -(1/2) \psi_1^2 \psi_2^2$, etc. The vanishing of the $\theta^2$ component,
is an automatic consequence of the fact that the integration of heavy scalars respects
supersymmetry. We checked that the $\theta^2$ component vanishes in all cases discussed,
which ensures the supersymmetry invariance of the solutions found.

%%%%%%%%%%%%%%%%%%%%%%%%%%%%%%%%%%%%%%%

\subsection{Yukawa couplings and generalized chiral constraints}
\label{sec:consy}

So far we discussed the impact of the Kahler potential on the polynomial
constraints.
A superpotential with additional, R-parity violating
coupling\footnote{The Higgs superfields constraints are different and
  will be discussed in Section \ref{sec:simplest} and
 Appendix \ref{sec:higgsinos}.} increases the order of
the monomial chiral constraints. Here we provide a generic example, with
\medskip
\bea
&& K \ = \ X^{\dagger} X +  Q^{\dagger} Q -
% \frac{m_X^2}{f^2}
\epsilon_x \ (X^{\dagger} X)^2 -
% \frac{m_q^2}{f^2}
\epsilon_q \ 
(Q^{\dagger} Q) (X^{\dagger} X) \ , \nonumber \\
&& W \ = \ \ f \ X \ + \frac{\lambda}{3} Q^3 \ . \label{consy1}
\eea
As before, one finds $m_X^2= 4 \epsilon_x f^2$ , $m_q^2=\epsilon_q
f^2$. 
In this case the integration of the heavy scalars leads to low-energy fields of the form
\bea
&& X  = 
d_{11}\,\psi_X\psi_X+
d_{12}\, \psi_q \psi_q
+ d_{13}\,\psi_X\,\psi_q+
d_{14}\, (\overline F_q\,\overline\psi_X
-\overline F_X\,\overline \psi_q)^2\,
 + \sqrt 2 \ \theta\,  \psi_X +  \theta^2  F_X, \nonumber \\[4pt]
&& Q  = 
d_{21}\, \psi_X\psi_X+d_{22}\,\psi_q \psi_q+d_{23}\,\psi_X\psi_q
 + d_{24}\, (\overline F_q\,\overline\psi_X-\overline F_X\,\overline \psi_q)^2\,
+ \sqrt 2 \ \theta \ \psi_q\ + \theta^2  F_q \ . \qquad
\label{consy2}
\eea
where we do not write explicitly the (long) expressions for $d_{ij}$.
By Grassmann variable arguments one can check that in
this case we obtain quartic constraints
\medskip
\bea
X^4 \ = \ X^3 Q \ = \ X^2 Q^2 \ = \ X Q^3 \ = \ Q^4 \ = \ 0 \qquad
{\rm and} \qquad  X^2\not= 0, \,\,\, X\,Q\not=0 \ .
\label{consy3}
\eea

\medskip\noindent
If $\lambda$ vanishes, one recovers the
conditions in (\ref{nl1}), (\ref{nl2}).

%%%%%%%%%%%%%%%%%%%%%%%%%%%%%%%%%%%%%%%%

\section{Leading-order low-energy effective  Lagrangians}
\label{sec:leading}

In this section we continue the analysis of the effects of additional
couplings in the superpotential on  the expression 
of the sgoldstino field, which lead earlier to modified superfield constraints.  
We study the goldstino couplings for  arbitrary Kahler potentials with an 
R-parity violating superpotential.
In the heavy scalars limit, non-universal couplings vanish and
the universal  couplings of the  goldstino are recovered, 
compatible with the universal  constraints.

%%%%%%%%%%%%%%%%%%%%%%%%%%%%%%%%%%%%%%%%%%

\subsection{Onshell Lagrangian with R-parity violating and goldstino couplings}
\label{sec:rp}

Let us consider $N$ superfields (quarks and/or leptons for MSSM) plus the goldstino
superfield $X$. We add the minimal high-energy Kahler potential needed to decouple all
scalars and add an R-parity violating coupling, denoted $\lambda_{ijk}$, symmetric in 
its indices. We consider
\medskip
\bea
&& K \ = \ X^{\dagger} X +  Q_i^{\dagger} Q_i - \epsilon_x \ (X^{\dagger} X)^2 -
\epsilon_i \ (Q_i^{\dagger} Q_i) (X^{\dagger} X) \ , \nonumber \\
&& W \ = \ \ f \ X \ + \frac{1}{3}\, \lambda_{ijk} Q_i Q_j Q_k \ , \label{rp1}
\eea

\medskip\noindent
where a sum over repeated indices is understood.
In this section and the following, to  avoid multiple indices for the
component fields of $Q_i=(\phi_{q_i},\psi_{q_i},F_{q_i})$ we shall use a
simplified notation $Q_i\equiv (\phi_i,\psi_i,F_i)$, where $i=\overline{1,N}$.
By integrating-out the $N+1$ massive scalars fields of mass
$m_X^2 = 4 \epsilon_x f^2$ and $m_i^2= \epsilon_i f^2$
 we obtain solutions of the generic form (\ref{consy2}).
Instead of writing the  exact solution, we use a simpler approach
and expand it in powers of $F_i/F_X$, as also done in Section~\ref{sec:consk}, 
eq.(\ref{taylor}). 
This is possible since onshell,  $F_X$   is much larger than $F_i$. 
In  the first order in this expansion, we find the solutions for 
sgoldstino and squarks/sleptons:
\medskip
\bea
&&\phi_X \ = \ \frac{\psi_X\psi_X }{2 F_X} \ +
\ \frac{4}{m_X^2 F_X} \lambda_{ijk}\, {\overline F}_i \,{\overline \psi}_j
 {\overline \psi}_k \ ,
\label{rp2} \label{scalars}
\\
&& \phi_{i} \ = \ \frac{\psi_X \psi_{i} }{F_X} \ - \ \frac{F_{i}}{2F_X^2} \psi_X\psi_X
\ - \ \frac{1}{m_i^2} \lambda_{ijk} \, {\overline \psi}_{j} {\overline \psi}_{k}
+ \frac{2}{m_i^2} \lambda_{ijk} {\overline F}_{j} \left( \frac{\overline \psi_X
\overline \psi_{k} }{\overline F_X} -
\frac{1}{m_k^2} \lambda_{klm} {\psi}_l {\psi}_m \right)  \ . \nonumber
\eea

\medskip\noindent
As before, with these sgoldstino and squarks fields, $X^2\not=0$ and $X\,Q_i\not=0$
and higher order constraints are respected instead.
If only interested in the four-fermion low-energy effective Lagrangian,
the auxiliary fields $F_i$ can be neglected, since they
contribute only starting with eight-fermion terms. As for
the auxiliary field $F_X$, its expression is
\medskip
\bea
 -F_X &=& f + \frac{1}{4 f^3}\, {\overline \psi_X}^2 \Box \psi_X^2 +
\frac{1}{f^3}\, (\overline \psi_X \overline \psi_i)\, \Box\, (\psi_X \psi_i)
+ \frac{1}{m_i^2 f} \, (\lambda_{ijk}
{\overline \psi}_j {\overline \psi}_k)\, ({\overline \lambda}_{imn} {\psi}^m {\psi}^n)
 \nonumber \\
& +&
 \frac{1}{m_i^2 f^2} \, ({\overline \lambda}_{ijk} {\psi}_j {\psi}_k)\, \Box\, (\psi_X \psi_i)
+ \frac{2}{m_i^2 f^2} \, ({\lambda}_{ijk} {\overline \psi}_j {\overline \psi}_k)\, \Box\,
 (\overline \psi_X \overline \psi_i)
\nonumber\\[3pt]
& +&
\frac{2}{m_i^4 f}\, ({\lambda}_{ijk} {\overline \psi}_j {\overline \psi}_k)\, \Box\,
 ({\overline \lambda}_{imn} {\psi}_m {\psi}_n) \ . \label{rp3}
\eea

\medskip\noindent
Using  this result, one finds  the on-shell low-energy Lagrangian,
up to four-fermion fields:
\medskip
\bea\label{opt}
 {\cal L}_{\rm eff} &=& - f^2 + {\cal L}_{\rm kin} (\psi_X,\psi_i) +
\frac{1}{4 f^2} {\overline \psi_X}^2 \Box \psi_X^2 +
\frac{1}{f^2} (\overline \psi_X \overline \psi_i) \Box (\psi_X \psi_i)
\label{rp4} \\
&+& \frac{1}{m_i^2}  (\lambda_{ijk} {\overline \psi}_j
{\overline \psi}_k) ({\overline \lambda}_{imn}
{\psi}_m {\psi}_n) + \frac{1}{m_i^2 f} \ ({\overline \lambda}_{ijk} {\psi}_j {\psi}_k)
\Box (\psi_X \psi_i)
\nonumber\\
& +&
\frac{1}{m_i^4} \ ({\lambda}_{ijk} {\overline \psi}_j {\overline \psi}_k)
\Box ({\overline \lambda}_{imn} {\psi}_m {\psi}_n) \ . \nonumber
\eea

\medskip\noindent
$\cL_{kin}$ denotes kinetic terms of $\psi_X,\psi_i$.
Note that, as expected, non-derivative terms involving the goldstino  canceled, by using
 the identity $\lambda_{ijk} (\psi_X \psi_i) ({\psi}_j {\psi}_k) =0 $. The first
line in (\ref{rp4}) gives the universal goldstino couplings  discussed extensively
in the literature, while the following lines give the model-dependent  couplings.
These are all couplings with up to four fermions.

In the absence of R-parity violating couplings, the fourth term in the rhs of (\ref{rp4})
 is the leading interaction between the goldstino and matter fermions and was discussed
in detail from a phenomenological viewpoint in \cite{bfz,bfmz}. In the presence of
R-parity couplings
$\lambda_{ijk}$, (\ref{rp4}) contains further interactions of phenomenological relevance.
Notice in particular the last term in the second line of (\ref{opt}), 
which can generate interactions of type $\psi_i\,\psi_j 
\rightarrow \psi_k \,\psi_X$, with a coupling $1/(m_i^2f)$. 
This coupling can be stronger then the $\cO(1/f^2)$ coupling of 
 two-goldstinos processes  
$\psi_i \psi_i \rightarrow \psi_X \psi_X + \ jet $ and
 $\psi_q \psi_q \rightarrow \psi_X \psi_X \gamma $ discussed in \cite{bfz,bfmz}.
It would then be useful to study the  phenomenological consequences of this
leading  process $\psi_i\,\psi_j \rightarrow \psi_k \,\psi_X$.
\footnote{For the Standard Model case, there are three R-parity
  violating couplings in the second line of
(\ref{rp4}), of the type $a_{ijk} u_i d_j \Box (d_k \psi_X)$, $b_{ijk} l_i q_j 
\Box (d_k \psi_X)$, $c_{ijk} l_i l_j \Box (e_k \psi_X)$ plus
permutations. The first one was analyzed in \cite{choi} where it was 
shown that it is severely constrained since it generates at tree-level
the proton decay mode $p \rightarrow \pi^+ {\bar \psi_X}$. The second can
generate a charged pion decay mode $\pi^+ \rightarrow e^+ \psi_X$. 
We thank G. Ferretti for discussions on this issue.}

We end this Section with a brief remark on the Goldstino couplings
in the low-energy effective Lagrangian versus those derived from
its microscopic version.
It seems that all possible goldstino couplings
obtained by starting from a microscopic  theory (as those 
in (\ref{rp4})) can be obtained starting instead from an effective Lagrangian
with any set of constraints that can eliminate the superpartners, to which one adds
appropriate sets of higher-dimensional and higher-derivative 
operators (suppressed by $m_i$ or $f$). 
To illustrate this idea, take the simplest set of constraints
of \cite{SK}, instead of those respected by the fields leading to  (\ref{rp4}).
One then notices that the Lagrangian (\ref{rp4}) can also be obtained
by starting from a low-energy Lagrangian containing the constrained
superfields $X^2= X Q_i = 0$, of the type 
\medskip\begin{multline}\label{ks1}
{\cal L}  = \int d^4 \theta \,\,
\Big[ X^{\dagger} X + Q_i^{\dagger} Q_i \
+
\frac{1}{m_i^2}\, \lambda_{ijk}\lambda_{imn}\,
Q_i^{\dagger} Q_j^{\dagger} Q_m Q_n +
\frac{1}{m_i^2}\,\lambda_{ijk} \, Q_i  Q_j D^2 Q_k
\\
+
\frac{1}{m_i^4}\,\lambda_{ijk}\lambda_{imn}
 Q_i^{\dagger} Q_j^{\dagger} \Box (Q_m Q_n) \Big] \
+ \
\Big\{\int d^2 \theta f X + {\rm h.c.}\Big\} \ .
\end{multline}

\medskip\noindent 
However, notice that a low-energy Lagrangian of the type given in (\ref{ks1}),
 with arbitrary coefficients in front of higher-dimensional 
operators {\it does not} necessarily correspond to a   
(microscopic) UV Lagrangian, which may not even exist. 
So the most general  low-energy theories defined in an expansion 
$E/m_i,E/\sqrt{f}$ might not have a UV completion and
end up in the so-called ``swampland'' \cite{vafa},
while only a subset of the low-energy theories can be derived from
simple UV theories.

%%%%%%%%%%%%%%%%%%%%%%%%%%%%%%%%%%%%%%%%%%%%%%%%%%%%%%%%%%%%%%%%%

\subsection{General onshell Lagrangian with general 4-fermion terms}

The results of Section~\ref{sec:rp} can be extended to an arbitrary Kahler potential.
For simplicity, we start directly with the onshell formulation of the Lagrangian
obtained from~(\ref{cons2})
\medskip
\bea
\mathcal L  &=&
- \ K^J_I\,\Big(\,\partial_\mu \phi_J^\dagger \partial_\mu \phi^I
+
\frac{i}{2}\mathcal D_\mu\psi^I\sigma^\mu\overline\psi_J+h.c.\,\Big)
-
\big(K^{-1}\big)_I^J\, W_J\overline W^I\nn\\
&+&
\frac{1}{4}\, R^{KL}_{IJ}\,\psi^I\psi^J \overline\psi_K\overline\psi_L
-
\frac{1}{2}\,\Big(\, W_{IJ}-\Gamma^K_{IJ}\,W_K\,\Big)\,\,\psi^I\psi^J+h.c.
\eea

\medskip\noindent
where $\Gamma$ and $R$ are the connection and curvature of the Kahler metric.
Here the capital indices $I,J,K...$ take the value $0$ to label
the goldstino superfield components $X=(\phi_X,\psi_X,F_X)$ and 
the values $i, j, k, \dots$, to label the (quark/lepton) matter superfields of components
$Q_i=(\phi_i,\psi_i,F_i), \dots$, respectively, with $i=1 \cdots N$,
where $N$ is the number of the matter fields. Similarly, the derivatives of $W$
and $K$ wrt the scalar components (or their hermitian conjugates)
are labeled with lower (upper) indices, respectively,
in the standard notation: $K_j=\partial K/\partial\phi^j$,
$K^j=\partial K/\partial\phi^{\dagger}_j$,
$K_0=\partial K/\partial\phi_X$, $K^0=\partial K/\partial\phi_X^\dagger$, etc.

Let us expand the Lagrangian around the minimum (ground state), in normal coordinates
%\medskip
\begin{eqnarray}
K^J_I&=&\delta^J_I+R^{JK}_{IL}\,\phi^L\, \phi_K^\dagger+\dots \qquad ,
\qquad 
(K^{-1})^J_I= \delta^J_I-R^{JK}_{IL}\,\phi^L\,\phi_K^\dagger+\dots\nn\\[4pt]
\Gamma^K_{IJ}&=&(K^{-1})^K_L \, K^L_{IJ}=R^{KL}_{IJ}\,\phi^\dagger_L+\dots
\end{eqnarray}

\medskip\noindent
The superpotential that we consider here is similar to that used in Section~\ref{sec:rp}
\medskip
\beq\qquad
W=fX   %+\frac{1}{2}b_{ij}z^iz^k
+\frac{1}{3}\,\lambda_{ijk}\, Q^iQ^jQ^k \qquad \ , 
\eeq

\medskip\noindent
where the contribution of the goldstino field is taken linear, to break SUSY.
One finds
\bea
\mathcal L&=&
- \ \partial_\mu\phi_X^\dagger\,\partial_\mu \phi_X
- \ \partial_\mu\phi_i^\dagger\,\partial_\mu \phi^i
+\tfrac{1}{4}R^{ij}_{kl}\,\overline\psi_i\overline\psi_j\psi^k\psi^\ell\nn\\[5pt]
&&+\, R^{00}_{00}\,\bigl|f\phi_X+\tfrac{1}{2}\psi_X\psi_X\bigr|^2
+R^{0k}_{0\ell}\,(f\phi_k^\dagger+\overline\psi_k
\overline\psi_X)(f\phi^\ell+\psi^\ell\psi_X)\nn\\[5pt]
&&+\, \left[R^{k\ell }_{0i}\,\overline\psi_k\overline\psi_\ell(f\phi^i+\psi_X\psi^i)
+\tfrac{1}{2}\,R^{k\ell }_{00}\,\overline\psi_k
\overline\psi_\ell(f\phi_X+\tfrac{1}{2}\psi_X\psi_X)
\right.\nn\\[4pt]
&&\, \left.-\lambda_{kij}\,\phi^k
\,\psi^i\,\psi^j
+h.c.\right]+\dots \ \ \label{4f1}
\eea

\medskip\noindent
where all terms that do not contribute to four-fermion interactions are not shown.
It is then straightforward to obtain the zero momentum expressions for the 
sgoldstino $\phi_X$
and squarks (sleptons) $\phi^l$, up to the bilinear order in the fermions:
\medskip
\begin{eqnarray}
\phi_X &=&-\frac{1}{2\,f}\,\psi_X\psi_X +
\frac{f}{2m_X^2}R^{00}_{ij}\,\psi^i\psi^j \ , 
\nn\\
\phi^\ell&=&-\frac{1}{f}\,\psi_X \,\psi^\ell-(M^{-2})^\ell_k\,\,
\big(\,
\overline\lambda^{kij}\,\overline\psi_i\overline\psi_j
-\frac{f}{2}\,R^{0k}_{ij}\,\psi^i\psi^j\,
\big) \ . \label{scalarsols}
\end{eqnarray}

\medskip\noindent
The first term in each equation is the same universal fermion bilinear already
encountered in eq.~(\ref{scalars}). Here we have defined the scalar masses
$(M^2)^i_j=-f^2R^{0i}_{0j}$ and\footnote{We assume for simplicity
that there is no mixing induced by the soft mass squared matrix between the goldstino and
matter fields.
}
$m_X^2=-f^2R^{00}_{00}$.
Using this in the Lagrangian of eq.~(\ref{4f1}) one finds
 that all terms involving goldstinos
but no derivatives cancel, to give
\medskip
\begin{multline}
\mathcal L^{\rm 4-fermion}_{\rm Goldstino}=
+
\frac{1}{4f^2}\,\overline \psi_X^2\,\Box\, \psi_X^2
+
\frac{1}{f^2}\,\overline \psi_X\overline \psi_i\,\Box\, (\psi_X\psi^i)\\
-
\frac{1}{4m_X^2}\,R^{00}_{ij}\, \overline \psi_X^2\,\Box\,(\psi^i\psi^j)
+
(M^{-2})^k_\ell\,\overline \psi_X\overline \psi_k\,\Box\,
\big(f^{-1}\,\overline\lambda^{\ell ij}\,\overline\psi_i\overline\psi_j
-
\frac{1}{2}\,R^{0\ell}_{ij}\,\psi^i\psi^j\big)+h.c.
\label{4f2}
\end{multline}

\medskip\noindent
In the first line one identifies again the universal terms,
independent of UV physics. These terms originate from the first terms in the rhs of
eqs.~(\ref{scalarsols}). Unlike the operators in the second line in (\ref{4f2}), 
they do not involve the scalar fields masses (as one would naively
expect from integrating out heavy states). Moreover, the  various 
cancellations between the non-derivative terms leading to eq.~(\ref{4f2})
are also entirely due to the first terms in the rhs of eqs.~(\ref{scalarsols}).
These cancellations are not accidental, but are
due to the underlying non-linearly realized  supersymmetry.
A similar result is found if instead of integrating out superpartners,
one uses field redefinitions, in which the fields
are shifted by  the first terms in eq.~(\ref{scalarsols}).
Such shift induces similar cancellations and generates the universal terms in eq.~(\ref{4f2}).
In Ref.~\cite{Luty:1998np} it was noticed that this redefinition can be lifted to a
goldstino-dependent supersymmetry transformation that can be used to prove that
no derivative-free couplings of the goldstino can ever occur in the
universal couplings.

For completeness, let us list the terms involving only matter fields:
\medskip
\begin{multline}
\mathcal L^{\rm 4-fermion}_{\rm matter}
=\frac{1}{4}\,R_{ij}^{k\ell}\,\psi^i\psi^j\,\overline\psi_k\overline\psi_\ell
+
\frac{f^2}{4m_X^2}\,\big(\,R_{00}^{ij}\,\,\overline\psi_i\overline\psi_j\,\big)
\,\Big(1 + \frac{\Box}{m_X^2}\Big) \,\Big(R^{00}_{k\ell}\,\psi^k\psi^\ell\,\Big)
\\[5pt]
+
\Big(\lambda_{mk\ell}\,\psi^k\psi^\ell-\frac{f}{2}R_{0m}^{k\ell}\,
\overline\psi_k\overline\psi_\ell\Big)\,\Big( M^{-2} + M^{-4}\Box\Big)^m_n
\Big(\overline\lambda^{nij}\,\overline\psi_i\overline\psi_j-\frac{f}{2}\,R^{0n}_{ij}\,
\psi^i\psi^j\Big) \ . 
\end{multline}

\medskip\noindent
This concludes the discussion on  goldstino and matter couplings, that goldstino 
induces, with up to four-fermion fields, for an arbitrary Kahler and a 
polynomial superpotential.

%%%%%%%%%%%%%%%%%%%%%%%%%%%%%%%%%%%%%%%%%%%%%

\section{The SM  coupled to goldstino using
universal constraints}
\label{sec:simplest}

In this section we provide an application to the Standard Model (SM), to identify
the effects of coupling the goldstino to the SM fields, in particular to the 
Higgs sector, in a superfield language.  
To this purpose we use an effective Lagrangian approach, ``promote''
any SM field to a superfield  and impose the universal superfield constraints   
on both goldstino and the  SM fields  \cite{SK}, to decouple their superpartners.  
This means that  all matter  superfields  are  in non-linear supersymmetry  
representations. We employ the universal constraints,
 since as argued at the end of Section \ref{sec:rp},
one may indeed use these  (instead of their non-universal extension)
to write a low energy effective Lagrangian, as long as a detailed 
microscopic (UV) picture is not sought. The strategy we 
employ is to write down all the interactions that one should 
include in order to obtain the usual Standard Model couplings/masses, 
and then deduce the additional Goldstino couplings they imply. At 
the end of this section, we also comment 
on the most relevant additional couplings that one could expect beyond 
these ``universal'' terms.

To write the effective interactions allowed, one uses that 
various couplings can be eliminated by using the constraints
that decouple the superpartners. These are
\medskip
\begin{equation}
X^2 \ = \ X Q_i \ = \ Q_i Q_j Q_k \ = \ 0 \  \label{ks2}
\end{equation}

\medskip\noindent
and the field equations for the constrained superfields
\medskip
\begin{eqnarray}
\frac{1}{4} X {\overline D}^2 X^{\dagger} & = &
f X \quad , \quad
\frac{1}{4} Q_i Q_j {\overline D}^2 X^{\dagger} \ = \  f Q_i Q_j,
\nonumber \\[5pt]
 X {\overline D}^2 Q_i^{\dagger} & = & 0, \qquad  \qquad \quad
Q_j {\overline D}^2 Q_i^{\dagger} \ = \ 0 . \label{ks3}
\end{eqnarray}

\medskip\noindent
Eqs.(\ref{ks3}) can be obtained, for example, from the following
action with Lagrange multiplier chiral superfields $\chi_0, \chi_i$, 
to enforce the constraints (\ref{ks2}):
\medskip
\begin{equation}
K \ = \ X^{\dagger} X \ + \ Q_i^{\dagger} Q_i,
\quad  \quad
W \ = \ f X \ + \ \frac{1}{2} \,\chi_0 X^2 \ + \ \chi_i \,X Q_i \ . \label{ks4}
\end{equation}

\medskip\noindent
The field equations for all the fields are then
\medskip
\begin{eqnarray}
\frac{1}{4} {\overline D}^2 X^{\dagger} & = & f + \chi_0 X + \chi_i Q_i
\quad , \quad \frac{1}{4} {\overline D}^2 Q_i^{\dagger} \ = \ \chi_i X \ ,
\nonumber \\
 X^2 & = & X Q_i \ = \ 0 \ .\label{ks5}
\end{eqnarray}

\medskip\noindent
After appropriately combing eqs.(\ref{ks5}) one obtains
(\ref{ks2}), (\ref{ks3}).  Notice that further nontrivial
relations can be obtained by taking appropriate (superspace) derivatives of the
constraints and of the equations of motion, such as 
$XD_\alpha X=0$, $Q_iQ_j D^2X=XD^2Q_iQ_j$, etc. 
This ends the list of constraints for the goldstino and matter superfields 
$Q$ (for quarks), with  similar constraints for the leptons case ($L$).

Regarding the constraints for the Higgs multiplets,  one has two cases. 
The first case is to consider two  Higgs doublets, as in the MSSM, ``promoted'' 
to superfields and impose the constraints  in (\ref{nl06}) to decouple the higgsinos; 
in this case one ends up at low energy with a two Higgs doublet model (2HDM). 
The second case is to impose (\ref{nl06}) on one Higgs superfield,
while the other is set to zero; in this case on has exactly the 
SM case with one Higgs doublet.

Let us consider the first case of two Higgs multiplets.
With squarks/sleptons decoupled by the above superfield constraints, 
the low-energy theory consists of the SM plus two Higgs doublets.
One can then write a general effective Lagrangian, that can be separated into
a universal part (that does not bring additional parameters) 
and a non-universal one. The universal part is similar to that of the MSSM.
The result is
\medskip
\begin{eqnarray}
 {\cal L} &= &\int d^4 \theta \Big[
X^{\dagger} X\left(1-P(H_i,H_i^\dagger)\right) 
+ \sum_{i: H,Q,U,D,L,E}\!\!\!\! \Phi_i^{\dagger} e^{V_i}
\Phi_i   \Big] + 
\bigg\{ 
\int d^2 \,\theta \Big[
 f X\Big(1+ \tilde P (H_i, H_i^{\dagger})\Big)  
\nonumber \\
  &+ & Y_{ij}^u Q_i U_j H_2+Y_{ij}^d Q_i D_j H_1 + Y_{ij}^l L_i E_j H_1 +  \frac{1}{4}\,
\big(\,{\rm{tr}}\, W_a^{\alpha} W_{\alpha}^a,\big) +  \mu H_1 H_2  \Big] + {\rm h.c.} 
\bigg\},
\label{ks6}
\end{eqnarray}

\medskip\noindent
Note  that, unlike the MSSM case, here all superfields  other than $H_{1,2}$ satisfy the 
aforementioned constraints,  to ensure that their superpartners are 
decoupled.  $H_{1,2}$ are also constrained. 
The dimensionless functions $P$ and $\tilde P$ are 
 general  functions of the 
 constrained (chiral) superfields $H_i,H_i^{\dagger}$, and without loss 
of generality we can assume $P$ to be real. Due to the constraints
$X {\overline D}_{\dot \alpha} H_i^{\dagger} = 0$, ($i=1,2$), $X V ({H_i,H_i^{\dagger}}) =$
chiral, for any function $V$ \cite{SK}. 
Because of this, the superfield redefinition
\medskip
\begin{equation}
X\to \frac{X}{\sqrt{1-P}}
\end{equation}

\medskip\noindent
is holomorphic and leads, together with (\ref{ks6}), to a contribution $V_F$
to the scalar potential
\medskip
\begin{equation}
V_F=f^2\frac{(1+\tilde P)(1+\tilde P^\dagger)}{1-P}= f^2+f^2(P+\tilde P+\tilde P^\dagger)
+f^2|P+\tilde P|^2+\dots
\,.\label{pot}
\end{equation}

\medskip\noindent
The  functions $P,\tilde P$,  can be expanded as
\begin{eqnarray}
 f^2 P ({H_i,H_i^{\dagger}}) &=& m_i^2 |H_i|^2 + B_\mu  (H_1 H_2+ {\rm
   h.c.})
 + \mathcal O|H^4| \ , \nonumber\\[4pt]
f^2 \tilde P({H_i,H_i^{\dagger}}) &=& \tilde m_i^2 |H_i|^2 + \tilde B_\mu  H_1 H_2 
+\tilde B_\mu' H_1^{\dagger} H_2^{\dagger} 
+\mathcal O|H^4| \ . 
\end{eqnarray}

\medskip\noindent
Together with the D-term contribution to the  potential,
which is similar to that in the MSSM, and ignoring the quartic terms
in $P,\tilde P$, one obtains the Higgs potential
\medskip
\begin{eqnarray}
V &=&
(m_i^2+2\tilde m_i^2) \,\, \vert h_i\vert^2
%+ (m_2^2+2\tilde m_2^2)  \vert h_2\vert^2
+ ( [B_\mu+\tilde B_\mu+\tilde B_\mu']\,h_1.h_2 + {\rm h.c.})  \nonumber \\[5pt]
& +&  \frac{1}{f^2}\,\Big\vert (m_i^2+\tilde m_i^2)\,\vert h_i\vert^2
  %+(m_2^2+\tilde m_2^2)\,\vert h_2\vert^2+
+(B_\mu+\tilde B_\mu)\,h_1.h_2
   +(B_\mu+\tilde B_\mu') h_1^{\dagger} h_2^{\dagger}
   \Big\vert^2  \nonumber \\
& + & \frac{g_1^2+g_2^2}{8}\,\Big[\vert h_1\vert^2-
\vert h_2\vert^2\Big]^2
+ \frac{g_2^2}{2}\,\vert h_1^\dagger\,h_2\vert^2  \ . \label{ks8}
\end{eqnarray}

\medskip\noindent
The mass parameter $\mu$ in (\ref{ks6}) does not
contribute to the scalar potential but gives additional 
higgs-goldstino interactions \cite{SK},  that we discuss shortly.

Rather interestingly,  $V$ contains a peculiar quartic correction, similar 
to that pointed out in \footnote{In the case of  \cite{adgt} 
a similar quartic Higgs correction, which was a 
square of the MSSM Higgs soft term contribution, lead to a tree-level 
increase of the Higgs mass to the LEP2  bound, for a low $\sqrt f \sim 2$ to $7$ TeV, 
while at large $f$ the usual MSSM case is recovered. For further discussions on
the corrections to the higgs potential in similar set-ups,  see \cite{Antoniadis:2010nb}.}
``non-linear'' MSSM model of \cite{adgt}.
The difference between the present potential, which is that of 
a two-Higgs doublet  model extension of the SM and the model in 
\cite{adgt} is that in  (\ref{ks8}), the new quartic terms in the
second line contain four additional parameters, 
(compared to the parameters in the first line),
 while in ``non-linear MSSM'' \cite{adgt}, the new quartic terms 
did not bring in additional parameters.
While the present setup leads to a  restrictive case of a 2HDM potential 
(with four new quartic couplings instead of seven in a general 2HDM), 
it looses its predictive power for the Higgs mass, relative to 
the model \cite{adgt}. Ultimately, this difference is not too surprising,
and is due to the fact that
a supersymmetric theory with superpartners projected out by superfield constraints 
(as in \cite{adgt}) is not necessarily identical to that built in an effective 
approach, by promoting to (non-linear) superfields the SM spectrum with two 
additional Higgs doublets.

So far we neglected the quartic terms in $P,\tilde P$. If present, they
bring corrections to $V$:
\begin{equation}
\Delta V \ = \ \frac{a_i f^2}{\Lambda^4} |h_i|^4 \ + \ \cdots \ . \label{ks9}
\end{equation}
If $a_i \sim m_i^4 \Lambda^4/f^4$, such terms give contributions to the quartic
Higgs couplings similar to those in the second line of (\ref{ks8}), generated by $F_X$.
One obtains then a full, general two-Higgs doublet model 
scalar potential, as expected. The values of $a_i$ are model-dependent 
and reduce the predictive power of the model to that of usual 2HDM case.

Let us now comment on the leading (in $1/f$) interactions of the Higgs with the goldstino.
From our Lagrangian, eq.~(\ref{ks6}), one can identify two sources.
 The first one is the kinetic term for the nonlinear Higgs multiplet, 
which is always present in any theory and hence universal. 
Using (\ref{nl07}), it becomes
\medskip
\begin{equation}
- i \ \psi_{h,i} \sigma^m \ \partial_m {\overline \psi_{h,i}} \quad \rightarrow \quad
 - \frac{i}{f^2} \ \psi_X \ \partial_m {\bar h}_i \ \sigma^m \ \Box
\ ({\overline \psi_X} h_i )  \ ,  
\label{higgsino4}
\end{equation}

\medskip\noindent
where the equations of motions  and an integration by parts were used 
in the last step. 
The second source is the term $\mu H_1 H_2$, that does not
contribute to the scalar potential, but gives instead additional 
higgs-goldstino interactions \cite{SK}.
Interestingly enough, for energies $E < \mu$, the leading
higgs-goldstino couplings originate from this term:
\medskip
\begin{equation}
- \mu \ \psi_{h,1} \psi_{h,2} \quad \rightarrow \quad -\frac{\mu}{f^2}
\ {\bar \psi}_X {\bar \psi}_X \ \partial^m h_1  \ \partial_m h_2 \ . \label{ks10}
\end{equation}

\medskip\noindent
Therefore, depending on the energy regime, one or the other of interactions 
(\ref{higgsino4}), (\ref{ks10}), provide the dominant higgs-goldstino couplings. 
We should comment that the $\mu H_1 H_2$ term is actually not needed to 
reproduce any of the couplings and masses of the 2HDM, in particular, $\mu$ is
 not related to the mass of any physical particle. However, it does in principle
 give an interaction comparable or even dominant over the ``universal'' 
one in (\ref{higgsino4}).
The Lagrangian in (\ref{ks6}) together with (\ref{higgsino4}), (\ref{ks10}) provide the
leading couplings of the model. This ends the discussion of the case of two 
light Higgs doublets.

Let us consider now the second case for the Higgs multiplets, 
in which one of the two Higgs doublets is heavy and is eliminated by 
imposing the superfield constraint $H_1=0$. In this
case, the Higgs potential can be obtained by setting $h_1=0$
in eq.~(\ref{ks8}). 
The Yukawa couplings for the up-quarks are
the usual ones, while the ones for the down-quarks and the
leptons have to be constructed in a different manner; 
they can instead arise from the Kahler operator
\begin{equation}
\int d^4 \theta \ \frac{1}{f} \ ( Y_{ij}^d \ X^{\dagger} Q_i D_j
H_2^{\dagger} \ + \  Y_{ij}^l \ X^{\dagger}  L_i E_j
H_2^{\dagger})  \ .  \label{ks010}
\end{equation} 

\medskip\noindent
Interestingly, the interactions of the fermions with the goldstino resulting 
from the two types of Yukawas will be the same. Using eqns.~(\ref{nl3}) and 
(\ref{nl07}) we see that they are given by
\begin{equation}
\frac{1}{f^2}\, (\psi_X\sigma^\mu\overline \psi_X)\, \left[Y^d_{ij}\,
 (q^i_L\bar d^j_R )\,(\partial_\mu h^\dagger)
+
Y^u_{ij}\, (q^i_L\bar u^j_R )\,(\partial_\mu h)+{\rm h.c.}
\right]
\end{equation}
These couplings are again universal for any theory reproducing the SM at low 
energy. They show a similar suppression ($\sim 1/f^2$) as the other universal 
interactions, resulting from kinetic terms, that we derived in eqns.~(\ref{opt}) 
and (\ref{4f2}).

Finally, let us comment on some additional couplings that one can write
in this approach that can be  of some interest\footnote{As with the $\mu$ 
term discussed above, the interactions here are not necessary to reproduce any SM couplings.}.
The lowest dimensional operators that one could imagine are dimension-four and 
dimension-five R-parity violating operators.
However, due to the constraints (\ref{nl60}), all these R-parity violating 
operators are vanishing
%\medskip
\begin{equation}
U_i D_j D_k \ = \ L_i Q_j D_k \ = \ L_i L_j E_k \ = \ Q_i Q_j Q_k L_m \ = \ 0 \ .
\label{ks11}
\end{equation}
The most dangerous proton-decay and flavor-violating operators are thus absent, 
and one does not need to impose any R-parity for them.
However, dangerous proton decay and flavour-violating operators appear in other
non-universal, model-dependent couplings in the theory. In particular, 
one can have couplings of the type
\begin{equation}
\int d^2 \theta \ \frac{1}{\Lambda^2}( a_{ijk} U_i D_j \Box D_k + b_{ijk} L Q_j
 \Box D_k + c_{ijk}  L_i L_j \Box E_k + {\rm permutations}) \ ,  \label{ks12}
\end{equation}
where the permutations are over the position of the box
operator. These are the same operators that were mentioned in
footnote 6 at the end of Section \ref{sec:rp}. It is likely that one thus 
still needs to impose R-parity to forbid these terms.

\section{Conclusions}

In this work we studied the couplings of the goldstino to matter fields
in a supersymmetric formalism with superfields endowed with constraints.
There are two classes of constraints that can be present. 
The first of these is now well-known, and
describes  universal constraints  that are independent of the superpartner masses
and of the UV details of the theory. This class describes, in an elegant way and
using a superfield formalism, the goldstino action and its universal
couplings to matter (super)fields, suppressed by inverse powers of $\sqrt f$.
In this paper we identified a second class of superfield constraints,
which are higher order polynomials in superfields. 
These  constraints are non-universal, i.e. they
help one to identify the non-universal couplings  of the goldstino.
These constraints recover the former, universal class
in the formal limit of infinite masses for superpartners.

An interesting aspect of the non-universal constraints is that their 
general solution (for sgoldstino and squarks/sleptons)  is usually a 
function of some arbitrary parameters, that can have similar values for different  
UV completions. For example,  in the case of cubic constraints the  solutions depended 
on two such  parameters.  In specific microscopic models,
these parameters can be expressed in terms of  the UV details,  such as UV
coefficients of the effective operators in the microscopic
action and also on the auxiliary fields or the fermionic components.
It may then be possible that  the class of generalised constraints
could capture aspects of UV physics (even in non-perturbative regime),
without the need for a detailed description of the UV physics.
The constraints we found could be used to describe interactions of the goldstino to
matter even in the limit where the usual effective expansion (in $E/m_i$) is not
rapidly convergent.
Finally, it should be mentioned that the link of the superfield constraints to
microscopic models is valid only at the leading order in a
light-fields derivative expansion. Consequently sub-leading terms in
the low-energy action are not easily mapped into terms in a
microscopic theory (see Appendix).

We computed the onshell Lagrangian with R-parity violating
and goldstino couplings for a minimal Kahler potential that 
enforces that all superpartners be massive enough and decouple at low energy.
The results were then extended to a general Kahler potential,
for which the Lagrangian of the goldstino and its couplings to matter fields with
up to four-fermions were computed. These couplings are suppressed by (positive)
powers of the supersymmetry breaking scale ($\sqrt f$) for the case of
universal couplings,  and by the superpartner masses (sgoldstino, squarks, sleptons) 
for the case of non-universal couplings. Finally,
the Lagrangian with up to  four-fermions involving only matter fields 
that was induced by decoupling the superpartners including the sgoldstino, 
was also computed, with the couplings suppressed by the curvature 
of the Kahler potential (of matter fields) or by the sgoldstino mass.

To write the low energy effective Lagrangian  of goldstino coupled 
to the SM matter fields,  our results suggest that one could in 
principle use either set of superfield constraints (minimal or non-minimal), 
provided one adds an appropriate number of 
higher-dimension and higher-derivative operators,  with coefficients 
that have to be matched to a microscopic (UV) Lagrangian. 
From this
point of  view the simpler constraints in \cite{SK} can be more practical compared to the generalized
constraints. However, the fact that the former follow only from very simple UV theories make
the connection between UV and IR theories more difficult. 
One negative outcome is that the dimension of an operator
is not enough to determine its low-energy relevance.

Ignoring this potential problem, we initiated the use of the simplest set of
constraints to couple the Standard Model to a light goldstino. 
In an appropriate two-higgs doublet SM extension that was supersymmetrized
non-linearly by constraints, we wrote the lowest order terms, in particular 
the Higgs potential.  This potential includes additional quartic higgs terms,
 similar to those investigated in \cite{adgt}, containing however 
four additional couplings (parameters). This has an impact on the 
predictive power of the model relative to the case discussed in \cite{adgt}.
Finally, we argued that the leading higgs-goldstino couplings could 
originate from the superpotential term $\mu H_1 H_2$. This interaction 
is non-universal in the sense that it does not contribute any interactions
 to the 2HDM besides couplings to goldstinos, and hence it is not needed for
 the parametrization of the theory.
This is in contrast to the  Kahler term $H_i^\dagger H_i$, which
generates Higgs kinetic terms as well as  goldstino interactions at the same time. 
Other, important, non-universal goldstino couplings arise from R parity violating
 operators that could lead to proton decay. Although the leading ones vanish
 due to the constraints, the sub-leading ones can be present and probably need
 to be forbidden by imposing R parity.

\section*{Acknowledgments}

{We thank I. Antoniadis, G. Ferretti, Z. Komargodski, C. Petersson, A. Romagnoni,
P. Tziveloglou and F. Zwirner for interesting discussions.
This work was partially supported by the European ERC Advanced
Grant 226371 MassTeV, by the CNRS PICS Nos. 3747 and 4172, by the European
Initial Training Network PITN-GA-2009-237920 and by the Agence Nationale de
la Recherche.}

\section*{Appendix}
\begin{appendix}

\renewcommand{\theequation}{A.\arabic{equation}}
\setcounter{equation}{0}  
\section{The effects of derivatives of light fields on the constraints.}
\label{lightderivs}

In deriving our various matter constraints from a microscopic Lagrangian,
we neglected the effects of the derivatives of the light fields (quarks and leptons). 
Including these, generates further terms in the equations for the scalars 
that can invalidate the simple, lowest order superfield constraints. 
This is true in particular for the simple case of Section~\ref{sec:consk}
with only $\epsilon_1, \epsilon_3$  non-zero and $\epsilon_2=\epsilon_4=0$. In this
case  $X^2=X Q =0$ as discussed in the text and in \cite{SK}. 
The microscopic action also contains terms with derivatives of the light 
(fermionic) fields of the type
\begin{eqnarray}
 \delta {\cal L} & = & - (1 - 4 \epsilon_1 |\phi_X|^2 - \epsilon_3 |\phi_q|^2)
 \ \psi_X i \sigma^m \partial_m {\bar \psi_X}
-  (1 - \epsilon_3 |\phi_X|^2) \ \psi_q i \sigma^m \partial_m {\bar
  \psi_q} \ \nonumber \\
 &+& \ ( \ \epsilon_3 {\bar \phi}_X \phi_q \psi_X i \sigma^m \partial_m {\bar
  \psi}_q + {\rm h.c.} \ ) \ . \label{consk10}
\end{eqnarray}
After including the effects of these terms when integrating out  the heavy scalars 
$\phi_X, \phi_q$ via their equations of motion, we find that the constraints $X^2=X Q =0$
are no longer satisfied. However, these terms add only higher-order corrections
(in derivatives and in $1/f$), which change the low-energy Lagrangian only at a
higher order in the number of fermions and derivatives.

%%%%%%%%%%%%%

\section{Heavy gauginos and constrained vector superfields}
\label{sec:consvector}
\renewcommand{\theequation}{B.\arabic{equation}}
\setcounter{equation}{0}

In Section~\ref{sec:constrained} we discussed the integrating out of 
massive scalar fields and the associated non-universal constraints. 
Here we consider the case of decoupling the heavy fermions instead.

To begin with, consider the case of decoupling the massive gauginos.
We provide a simple microscopic realization of models with heavy gauginos
that we integrate out and then write down the low-energy constrained vector 
multiplet. We then compare the result to the constraint in (\ref{nl5}) \cite{SK},
that is supposed to decouple the massive gauginos.
The simplest UV Lagrangian providing large masses to the sgoldstino 
and gaugino is
\medskip
\begin{eqnarray}
{\cal L} & = & \int d^4 \theta \ \left[ X^{\dagger} X  \ - \ \epsilon
\,(X^{\dagger} X)^2 \right] \nonumber \\
&+& \bigg\{
\int d^2 \theta \ \Big( f X \ + \ \frac{1}{4} W^{\alpha} W_{\alpha} \
+ \ \frac{M}{f} \ X \ W^{\alpha} W_{\alpha}
\Big) + {\rm h.c.} \bigg\} \ , \label{vector1}
\end{eqnarray}

\medskip\noindent
where $M$ is the gaugino mass and the discussion applies to both
the Abelian and the non-Abelian cases. The zero-momentum equation of motion for
the sgoldstino $X$ gives the same solution as in (\ref{nl1}), (\ref{nl01}). 
However, if we keep the  momenta of the light fields,
similarly to the heavy scalar case discussed in Appendix \ref{lightderivs}, 
there are corrections  of the type
\begin{equation}
\phi_X \ = \ \frac{\psi_X \psi_X}{2 F_X} \ + 
\ \frac{M}{4 \epsilon f |F_X|^2} {\bar L}_\alpha^{\beta} {\bar L}_{\beta}^{\alpha} \ ,
\end{equation}
where $L_{\alpha}^{\beta} = \delta_{\alpha}^{\beta} D 
- i \sigma_{\alpha}^{mn, \beta} F_{mn}$. 
However, these corrections change the constraint
$X^2=0$ and the discussion below only 
by terms of high-order in the low-energy Lagrangian, 
that we can neglect in what follows.
Further, let us write the  equation for the gaugino $\lambda$, by keeping for the
time being the momentum-dependent terms. We find
\medskip
\begin{equation}
- i \sigma^m \partial_m {\overline \lambda} + \frac{M}{f}
\left[ - 2 F_X \lambda + i \sqrt{2} (D-i \sigma^{mn} F_{mn}) \psi_X -
2i \phi_X \sigma^m \partial_m {\overline \lambda} - 2i
\partial_m  ({\overline \phi}_X {\overline \lambda} {\overline \sigma}^m)
\right] \ = \ 0 \ . \label{vector2}
\end{equation}
The zero-momentum (in the) gaugino equation has the solution
\medskip
\begin{equation}
\lambda \ = \ \frac{i}{\sqrt{2}F_X} ( D-i \sigma^{mn} F_{mn} ) \ \psi_X \ .
\label{vector3}
\end{equation}

\medskip\noindent
The corresponding (supersymmetric) gauge field strength $W_{\alpha}$ is then given by
\medskip
\begin{equation}
W_{\alpha} \ = \ \frac{1}{\sqrt{2}F_X} (\delta_{\alpha}^{\beta} D
-i \sigma_{\alpha}^{mn,\beta} F_{mn}) \psi_{X,\beta}
+ (\delta_{\alpha}^{\beta} D-i \sigma_{\alpha}^{mn,\beta} F_{mn}) 
\theta_{\beta} + \theta^2  (\sigma^m \partial_m {\overline \lambda})_{\alpha}
 \ \label{vector03}
\end{equation}

\medskip\noindent
and satisfies the following equations (by using $\phi_X^0$ with the constraint $X^2=0$)
\medskip
\begin{equation}
X \ W_{\alpha} \ = \ \frac{\psi_X\psi_X}{2 F_X} \
(\sigma^m \partial_m {\overline \lambda})_{\alpha} \ \theta^2, \quad  \quad
X \ W^{\alpha} \ W_{\alpha} \ = \ 0 \ . \label{vector4}
\end{equation}

\medskip\noindent
The first constraint in (\ref{vector4}) is however not invariant under 
supersymmetry transformations. At first sight this seems
puzzling, since the second eq. in (\ref{vector4}) is manifestly supersymmetric. 
The explanation is that the 
general solution of $X \ W^{\alpha} \ W_{\alpha}$, although it 
contains (\ref{vector3}), it is in fact more general
\begin{equation}
\lambda \ = \ \frac{i}{\sqrt{2}F_X} ( D-i \sigma^{mn} F_{mn} ) 
\ \psi_X \ + \ \chi_{\alpha} \Psi_X \Psi_X \ ,
\label{vector030}
\end{equation}

\medskip\noindent
where $\chi_{\alpha}$ is an arbitrary Weyl fermion. The situation 
here is similar to that described for the heavy scalars in
eqs. (\ref{consk4}) and (\ref{consk5}). Indeed, the solutions of the
cubic constraints depend on the free 
parameters $c_{1,2}$, that have however nontrivial 
supersymmetry transformations.

Let us compare these results with (\ref{nl5}), (\ref{nl6}). The 
difference between them
is that in the latter case the  gauginos are the solution \cite{SK} of 
the implicit equation
\medskip
\begin{equation}
\lambda \ = \ \frac{i}{\sqrt{2}F_X} ( D-i \sigma^{mn} F_{mn}) \,\psi_X \,
- \ i \frac{\psi_X\psi_X}{2 F_X^2} \sigma^m \partial_m {\overline \lambda} \ ,
\label{vector5}
\end{equation}

\medskip\noindent
instead of  the solution in (\ref{vector3}). Notice that this corresponds to the 
spinor $\chi_{\alpha}$ in (\ref{vector030})
being equal to $\chi_{\alpha} = - i \sigma^m \partial_m {\bar \lambda}/ 2 F_X^2$.
The solution in (\ref{vector5}) can be obtained more easily by setting to zero 
the square bracket in (\ref{vector2}) (i.e. the terms proportional
to $M/f$), which corresponds to taking the limit $M \rightarrow \infty$
 in the gaugino field equation.
A similar self-consistent procedure in the field eqs for $\phi_X$ is however ambiguous,
since the corresponding field equation involves both the  sgoldstino and gaugino mass.
Notice however that the difference between (\ref{nl5}),(\ref{nl6}) and (\ref{vector3}) is 
of
higher-order in an $1/f$ expansion in the low-energy action. Therefore, in both cases the
 leading goldstino-gauge field interaction comes from the gaugino kinetic term by using
the common terms proportional to $\sigma^{mn} F_{mn} \ \psi_X $ in (\ref{vector3}).
We recover again the situation encountered earlier: different constraints,
the simplest with a unique solution (\ref{nl5}) 
and the higher-order one (\ref{vector4}) share the same leading 
term, which fixes the universal coupling of goldstino to gauge fields.

%%%%%%%%%%%%%%%%%%%%%%%%%%%%%%%%%%%%%%%%%%%%%%%%%%%%%%%%%%%%%%%%%%%%%%
\section{Heavy higgsinos.}
\label{sec:higgsinos}
\renewcommand{\theequation}{C.\arabic{equation}}
\setcounter{equation}{0}

In this section we consider the case of projecting out 
the massive Higgs superpartners (higgsinos).  
The standard example is the MSSM with its two Higgs 
multiplets $H_{1,2}$, in which the two higgsinos $\psi_{1,2}$ are massive;
this leads at low-energy to two nonlinear Higgs multiplets, that contain
as physical degrees of freedom only the complex scalars. 
To simplify the equations, in what follows  we consider only one chiral 
multiplet $H = (\phi_h,\psi_h,F_h)$ and decouple the corresponding fermion
$\psi_h$ by assuming it has a large mass $M$. The simplest Lagrangian 
giving a large mass to the fermion but not to the scalar Higgs $\phi_h$ is
\medskip
\begin{eqnarray}
 {\cal L} & = & \int d^4 \theta \ \Big[
 X^{\dagger} X + H^{\dagger} H - \ \epsilon\, (X^{\dagger} X)^2 +
\frac{M}{4f^2} X^{\dagger} X ( D^{\alpha} H D_{\alpha} H +
 {\overline D}_{\dot \alpha} H^{\dagger}
{\overline D}^{\dot \alpha} H^{\dagger}  ) \Big]
\nonumber \\
&+&  \bigg\{
\int d^2 \theta \ f X \ + \ {\rm h.c.}
\bigg\}, \label{higgsino1}
\end{eqnarray}

\medskip\noindent
The component Lagrangian at zero-momentum for the heavy fields $\phi_X, \psi_h$ is
\medskip
\begin{eqnarray}\label{higgsino2}
 {\cal L} & = &\!\!
(1 - 4 \epsilon\,|\phi_X|^2) |F_X|^2 + |F_h|^2 +
\bigg\{ {\overline F}_X \, (f + 2 \epsilon\, {\overline \phi}_X\, \psi_X\psi_X )
 + \frac{M}{f^2} \Big[
|\phi_X|^2 (F_h \Box h
 - \partial F_h\, \partial h)\nonumber\\
&  -&
 {\overline \phi}_X F_X (\partial h)^2 + \phi_X {\overline F}_X F_h^2
+ i F_h \partial_m h \psi_X \sigma^m {\overline \psi_X}
 - i F_X  \partial_m h \psi_h \sigma^m {\overline \psi_X}
\nonumber\\
&-& {\overline F}_X F_h \psi_X \psi_h + \frac{1}{2} \Big(
  |F_X|^2 + \frac{i}{2} \partial_m {\overline \psi_X} {\overline \sigma}^m \psi_X
 -\frac{i}{2} {\overline \psi_X}  {\overline \sigma}^m \partial_m  \psi_X
\Big) \psi_h \psi_h \Big]  \!+\! {\rm h.c.}\! \bigg\}
\end{eqnarray}

\medskip\noindent
The field equation for the higgsino $\psi_h$ and the auxiliary field
$F_h$ give
\medskip
\begin{equation}
\psi_h \ = \ \frac{i}{{\overline F}_X} \ \sigma^m {\overline \psi_X}
\ \partial_m \, h \  \quad , \quad 
F_h \ = \ - \frac{2M}{f^2} |\phi_X|^2 \Box h \ . \ 
\label{higgsino3}
\end{equation}
We therefore find the solution for the higgsino shown in eq.(\ref{nl07}).

When comparing our microscopic theory to 
the results of using the Higgs superfield constrained as in (\ref{nl06}),  (\ref{nl07})
one encounters a puzzle that is not yet clarified. 
This refers to the fact that there is no simple, clear 
interpretation in the microscopic theory of the fact that the auxiliary 
field $F_h$  in (\ref{nl07}) is not dynamical.

\end{appendix}

%\newpage


\begin{thebibliography}{99}


  %\cite{Volkov:1973ix}
\bibitem{Volkov:1973ix}
  D.~V.~Volkov, V.~P.~Akulov,
  ``Is the Neutrino a Goldstone Particle?,''
  Phys.\ Lett.\  B {\bf 46} (1973) 109.
  %%CITATION = PHLTA,B46,109;%%

\bibitem{cddfg}
%\cite{Fayet:1977vd}
%\bibitem{Fayet:1977vd}
  P.~Fayet,
  ``Mixing between Gravitational And Weak Interactions Through The Massive
  Gravitino,''
  Phys.\ Lett.\  B {\bf 70} (1977) 461.
  %%CITATION = PHLTA,B70,461;%%
%\cite{Fayet:1979qi}
%\bibitem{Fayet:1979qi}  P.~Fayet,
``Weak Interactions Of A Light Gravitino: A Lower Limit On The Gravitino Mass
From The Decay Psi $\to$ Gravitino Anti-Photino,''
  Phys.\ Lett.\  B {\bf 84} (1979) 421.
  %%CITATION = PHLTA,B84,421;%%
%\cite{Fayet:1979yb}
%\bibitem{Fayet:1979yb} P.~Fayet,
``Scattering Cross-Sections Of The Photino And The Goldstino (Gravitino) On
Matter,'' Phys.\ Lett.\  B {\bf 86} (1979) 272.
%%CITATION = PHLTA,B86,272;%%
%\cite{Casalbuoni:1988kv}
R.~Casalbuoni, S.~De Curtis, D.~Dominici, F.~Feruglio and R.~Gatto,
``A Gravitino - Goldstino High-Energy Equivalence Theorem,''
Phys.\ Lett.\ B {\bf 215} (1988) 313
%%CITATION = PHLTA,B215,313;%%
% \bibitem{Casalbuoni:1988qd}
% R.~Casalbuoni, S.~De Curtis, D.~Dominici, F.~Feruglio, R.~Gatto,
``High-Energy Equivalence Theorem In Spontaneously Broken Supergravity,''
Phys.\ Rev.\ D {\bf 39} (1989) 2281.
%%CITATION = PHRVA,D39,2281;%%

%\cite{Ivanov:1977my}
\bibitem{Ivanov:1977my}
  E.~A.~Ivanov, A.~A.~Kapustnikov,
``Relation Between Linear And Nonlinear Realizations Of Supersymmetry,''
  %%CITATION = JINR-E2-10765;%%
%\cite{Ivanov:1978mx}
%\bibitem{Ivanov:1978mx}
%  E.~A.~Ivanov and A.~A.~Kapustnikov,
``General Relationship Between Linear And Nonlinear Realizations Of
Supersymmetry,''
  J.\ Phys.\ A  {\bf 11} (1978) 2375.
  %%CITATION = JPAGB,A11,2375;%%
%\cite{Ivanov:1982bpa}
%\bibitem{Ivanov:1982bpa}
%  E.~A.~Ivanov and A.~A.~Kapustnikov,
``The Nonlinear Realization Structure Of Models With Spontaneously Broken
Supersymmetry,''
  J.\ Phys.\ G {\bf 8} (1982) 167.
  %%CITATION = JPHGB,G8,167;%%


%\cite{Rocek:1978nb}
\bibitem{Rocek:1978nb}
  M.~Rocek,
  ``Linearizing The Volkov-Akulov Model,''
  Phys.\ Rev.\ Lett.\  {\bf 41} (1978) 451.
  %%CITATION = PRLTA,41,451;%%

%\cite{Lindstrom:1979kq}
\bibitem{Lindstrom:1979kq}
  U.~Lindstrom, M.~Rocek,
  ``Constrained Local Superfields,''
  Phys.\ Rev.\  D {\bf 19} (1979) 2300.
  %%CITATION = PHRVA,D19,2300;%%


%\cite{Samuel:1982uh}
\bibitem{Samuel:1982uh}
S.~Samuel and J.~Wess,
``A Superfield Formulation Of The Nonlinear Realization Of Supersymmetry And
Its Coupling To Supergravity,''
Nucl.\ Phys.\ B {\bf 221} (1983) 153.
%%CITATION = NUPHA,B221,153;%%

%\cite{Casalbuoni:1988xh}
\bibitem{Casalbuoni:1988xh}
  R.~Casalbuoni, S.~De Curtis, D.~Dominici, F.~Feruglio and R.~Gatto,
  ``Nonlinear realization of supersymmetry algebra from supersymmetric
  constraint''
  Phys.\ Lett.\  B {\bf 220} (1989) 569.
  %%CITATION = PHLTA,B220,569;%%


%\cite{Clark:1996aw}
\bibitem{Clark:1996aw}
T.~E.~Clark and S.~T.~Love,
``Goldstino couplings to matter,''
Phys.\ Rev.\ D {\bf 54} (1996) 5723
[arXiv:hep-ph/9608243].
%%CITATION = HEP-PH 9608243;%%


\bibitem{Clark:1998aa}
T.~E.~Clark, T.~Lee, S.~T.~Love and G.~Wu,
``On the interactions of light gravitinos,''
Phys.\ Rev.\ D {\bf 57} (1998) 5912
[arXiv:hep-ph/9712353].
%%CITATION = HEP-PH 9712353;%%

\bibitem{Brignole:1997pe} A.~Brignole, F.~Feruglio and F.~Zwirner,
``On the effective interactions of a light gravitino with matter
fermions,'' JHEP {\bf 9711} (1997) 001
[arXiv:hep-th/9709111].

\bibitem{bfz}
 A.~Brignole, F.~Feruglio, F.~Zwirner,
 ``Signals of a superlight gravitino at e+ e- colliders when 
the other superparticles are heavy,''
  Nucl.\ Phys.\  {\bf B516 } (1998)  13-28.
  [hep-ph/9711516].

\bibitem{bfmz}
 A.~Brignole, F.~Feruglio, M.~L.~Mangano, F.~Zwirner,
 ``Signals of a superlight gravitino at hadron colliders when the other 
superparticles are heavy,''
  Nucl.\ Phys.\  {\bf B526 } (1998)  136-152.
  [hep-ph/9801329].


\bibitem{Luty:1998np}
M.~A.~Luty and E.~Ponton,
``Effective Lagrangians and light gravitino phenomenology,''
Phys.\ Rev.\ D {\bf 57} (1998) 4167
hep-ph/9706268,v3 [revised version of Phys.\ Rev.\ D
{\bf 57} (1998) 4167].
%%CITATION = HEP-PH 9706268;%%


%\cite{Antoniadis:2004se}
  \bibitem{Antoniadis:2004uk}
  I.~Antoniadis and M.~Tuckmantel,
  ``Non-linear supersymmetry and intersecting D-branes,''
  Nucl.\ Phys.\  B {\bf 697} (2004) 3
  [arXiv:hep-th/0406010];
  %%CITATION = NUPHA,B697,3;%%
  I.~Antoniadis, M.~Tuckmantel and F.~Zwirner,
``Phenomenology of a leptonic goldstino and invisible Higgs boson decays,''
  Nucl.\ Phys.\  B {\bf 707} (2005) 215
  [arXiv:hep-ph/0410165].
  %%CITATION = NUPHA,B707,215;%%
%\cite{Antoniadis:2004uk}


%\cite{Brignole:2003cm}
\bibitem{Brignole:2003cm}
A.~Brignole, J.~A.~Casas, J.~R.~Espinosa and I.~Navarro,
``Low-scale supersymmetry breaking: Effective description,
electroweak breaking and phenomenology,''
  Nucl.\ Phys.\  B {\bf 666} (2003) 105
  [arXiv:hep-ph/0301121].
  %%CITATION = NUPHA,B666,105;%%

%\cite{Komargodski:2009rz}
\bibitem{SK}
Z.~Komargodski and N.~Seiberg,
``From Linear SUSY to Constrained Superfields,''
JHEP {\bf 0909} (2009) 066 [arXiv:0907.2441 [hep-th]].
%%CITATION = JHEPA,0909,066;%%

\bibitem{AlvarezGaume:2010rt}
  L.~Alvarez-Gaume, C.~Gomez and R.~Jimenez,
  ``Minimal Inflation,''
  Phys.\ Lett.\  B {\bf 690} (2010) 68
  [arXiv:1001.0010 [hep-th]].
  %%CITATION = PHLTA,B690,68;%%

%\cite{Cheung:2011jq}
\bibitem{Cheung:2011jq}
 C.~Cheung, F.~D'Eramo, J.~Thaler,
 ``The Spectrum of Goldstini and Modulini,''    
[arXiv:1104.2600 [hep-ph]].
%\cite{Cheung:2010mc}
%\bibitem{Cheung:2010mc}
 C.~Cheung, Y.~Nomura, J.~Thaler,
 ``Goldstini,''
 JHEP {\bf 1003 } (2010)  073.
 [arXiv:1002.1967 [hep-ph]].

%\cite{Craig:2010yf}
 \bibitem{Craig:2010yf}
 N.~Craig, J.~March-Russell, M.~McCullough,
 ``The Goldstini Variations,''
 JHEP {\bf 1010 } (2010)  095.
 [arXiv:1007.1239 [hep-ph]].

%\cite{Argurio:2011hs}
\bibitem{Argurio:2011hs}
 R.~Argurio, Z.~Komargodski and A.~Mariotti,
 ``Pseudo-Goldstini in Field Theory,''
 [arXiv:1102.2386 [hep-th]].

\bibitem{dine}
  M.~Dine, G.~Festuccia, Z.~Komargodski,
  ``A Bound on the Superpotential,''
  JHEP {\bf 1003 } (2010)  011.
  [arXiv:0910.2527 [hep-th]].

\bibitem{adgt}
%\cite{Antoniadis:2010hs}
%\bibitem{Antoniadis:2010hs}
  I.~Antoniadis, E.~Dudas, D.~M.~Ghilencea and P.~Tziveloglou,
  ``Non-linear MSSM,''
  Nucl.\ Phys.\  B {\bf 841} (2010) 157
  [arXiv:1006.1662 [hep-ph]].
  %%CITATION = NUPHA,B841,157;%%

\bibitem{WB}
J. Wess and J. Bagger, ``Supersymmetry and Supergravity'', second edition,
Princeton University Press, 1992~; P.~Binetruy, G.~Girardi, R.~Grimm,
``Supergravity couplings: A Geometric formulation,''
  Phys.\ Rept.\  {\bf 343 } (2001)  255-462.
  [hep-th/0005225].

\bibitem{zumino}
  B.~Zumino,
  ``Supersymmetry and Kahler Manifolds,''
  Phys.\ Lett.\  {\bf B87 } (1979)  203.

\bibitem{vafa}
  C.~Vafa,
  ``The String landscape and the swampland,''
  [hep-th/0509212].

\bibitem{choi}
 K.~Choi, E.~J.~Chun, J.~S.~Lee,
  ``Proton decay with a light gravitino or axino,''
  Phys.\ Rev.\  {\bf D55 } (1997)  3924-3926.
  [hep-ph/9611285].

%\cite{Antoniadis:2010nb}
\bibitem{Antoniadis:2010nb}
  I.~Antoniadis, E.~Dudas, D.~M.~Ghilencea, P.~Tziveloglou,
  ``Beyond the MSSM Higgs with d=6 effective operators,''
  Nucl.\ Phys.\  {\bf B848 } (2011)  1-32.
  [arXiv:1012.5310 [hep-ph]].
%%CITATION = NUPHA,B848,1;%%
%\cite{Antoniadis:2009rn}
% \bibitem{Antoniadis:2009rn}
%  I.~Antoniadis, E.~Dudas, D.~M.~Ghilencea, P.~Tziveloglou,
  ``MSSM Higgs with dimension-six operators,''
  Nucl.\ Phys.\  {\bf B831 } (2010)  133-161.
  [arXiv:0910.1100 [hep-ph]].
%%CITATION = NUPHA,B831,133;%%
%\cite{Antoniadis:2008es}
% \bibitem{Antoniadis:2008es}
%  I.~Antoniadis, E.~Dudas, D.~M.~Ghilencea, P.~Tziveloglou,
  ``MSSM with Dimension-five Operators (MSSM(5)),''
  Nucl.\ Phys.\  {\bf B808 } (2009)  155-184.
  [arXiv:0806.3778 [hep-ph]].
 %%CITATION = NUPHA,B808,155;%%
%\cite{Cassel:2009ps}
%\bibitem{Cassel:2009ps}
  S.~Cassel, D.~M.~Ghilencea, G.~G.~Ross,
``Fine tuning as an indication of physics beyond the MSSM,''
  Nucl.\ Phys.\  {\bf B825 } (2010)  203-221.
  [arXiv:0903.1115 [hep-ph]].
%%CITATION = NUPHA,B825,203;%%
%%CITATION = ARXIV:1101.4664;%%
%\cite{Carena:2009gx}
%\bibitem{Carena:2009gx}
 M.~Carena, K.~Kong, E.~Ponton, J.~Zurita,
 ``Supersymmetric Higgs Bosons and Beyond,''
 Phys.\ Rev.\  {\bf D81 } (2010)  015001.
 [arXiv:0909.5434 [hep-ph]].
 %%CITATION = PHRVA,D81,015001;%%
%\cite{Dine:2007xi}
% \bibitem{Dine:2007xi}
 M.~Dine, N.~Seiberg, S.~Thomas,
 ``Higgs physics as a window beyond the MSSM (BMSSM),''
 Phys.\ Rev.\  {\bf D76 } (2007)  095004.
 [arXiv:0707.0005 [hep-ph]].
\end{thebibliography}
\end{document}